\documentclass[a4paper,10pt]{article}
\usepackage[T1]{fontenc} 
\usepackage{fixltx2e} 
\usepackage{multirow}
\usepackage{epsfig}
\usepackage{rotating}
\usepackage{amsmath}
\usepackage{amssymb}
\usepackage{psboxit}
\usepackage[tight]{subfigure}
\usepackage{rotating}
\usepackage{euscript}
\usepackage{graphics}
\usepackage{txfonts}
\usepackage[usenames]{color}
\usepackage{graphicx}
\usepackage[colorlinks=false]{hyperref}

\input{hindex.mac}

\title{Categorizing Influential Authors \\ Using Penalty Areas}

\author{Antonis Sidiropoulos$^{1,3}$ \ \ Dimitrios Katsaros$^2$
\thanks{Corresponding author: Dimitrios Katsaros ({\tt dkatsar@inf.uth.gr})} 
\ \ Yannis Manolopoulos$^1$ \\ \\
$^1$Department of Informatics, \\ Aristotle University, Thessaloniki, Greece \\ \\
$^2$Department of Computer \& Communications Engineering, \\ University of Thessaly, Volos, Greece \\ \\ 
$^3$Department of Information Technology, \\ 
Alexander Technological Educational Institute of Thessaloniki, \\ Thessaloniki, Greece \\ \\
\tt \{asidirop,manolopo\}@csd.auth.gr, dkatsar@inf.uth.gr
}

\begin{document}
\maketitle

\begin{abstract}
The concept of \hi has been proposed to easily assess a researcher's performance with a single two-dimensional number. 
However, by using only this single number, we lose significant information about the distribution of the number of citations per article of an author's publication list. 
Two authors with the same \hi may have totally different distributions of the number of citations per article. 
One may have a very long "tail" in the citation curve, i.e. he may have published a great number of articles, which did not receive relatively many citations. 
Another researcher may have a short tail, i.e. almost all his publications got a relatively large number of citations. 
In this article, we study an author's citation curve and we define some areas appearing in this curve.
These areas are used to further evaluate authors' research performance from quantitative and qualitative point of view. 
We call these areas as "penalty" ones, since the greater they are, the more an author's performance is penalized. 
Moreover, we use these areas to establish new metrics aiming at categorizing researchers in two distinct categories: "influential" ones vs. "mass producers".
\end{abstract}

\section{Introduction}
\label{sec:intro}

\begin{figure}[ht]
\centerline{\psfig{figure=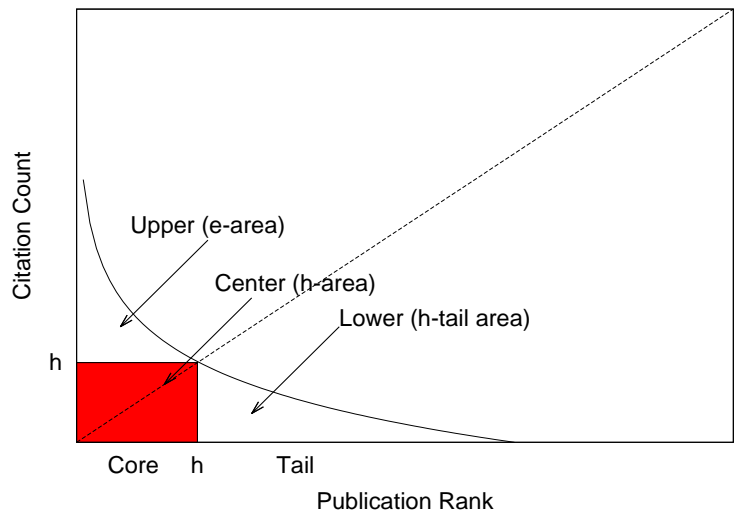,clip=,width=4in}}
\caption{Citation curve with upper, core, and lower areas.}
\label{example_plain_fig}
\end{figure}

The \hi has been a well honored concept since it was proposed by Jorge Hirsh in 2005 \cite{hirsch05}. 
Several variations have proposed in the literature \cite{wiki}, which have been implemented in commercial and free software, such as Matlab and Publish or Perish \cite{pubperish2010}. 
Recently, there have appeared several studies focusing on specific parts of the citation curve.
As discussed in \cite{Rosenberg11}, the citation curve is divided in three areas (see Figure~\ref{example_plain_fig}).
The first area is a square of size $h$ and is called {\it core area}.
This area is depicted by grey color in Figure~\ref{example_plain_fig} and, apparently, includes $h^2$ citations. 
The area that lies to the right of the core area is called {\it tail area} or {\it lower area}, whereas the area above the core area may be called {\it head area} or {\it upper area} or $e^2$ area \cite{Zhang09}.

Apparently, the publications in the tail area do not get many citations (certainly, less than the respective \hi).
On the other hand, the upper area comprises of citations to papers, which contribute to the calculation of the \hi; however, the numbers of citations to these papers are larger than $h$ and in some sense they are "wasted". 
For this reason, this area is also called {\it excess area}.
To overcome the specific deficiency of \hi, the {\it $g$-index} was proposed by Leo Egghe \cite{egghe06,egghe06b}.

In article \cite{Rosenberg11} the scientists with many citations in the upper area and a few citations in the tail area are characterized as {\it perfectionists}. 
These are the scientists, which have authored publications mostly of high impact.
Authors with few citations in the upper area and many citations in the tail area are characterized as {\it mass producers}, since they have publications mostly of relatively low impact. 
Those with moderate figures of citations in the upper and tail areas are named {\it prolific} (i.e. they have produced an abundance of influential papers). 
The tail and the excess area give significant information about the researcher's performance. 
The Tail to Core ratio has been studied in \cite{Rousseau09} but the information of the tail length is lost.
In the present paper, we try to devise a methodology and an easy criterion to categorize a scientist in one of two distinct categories: either an author is a "mass producer" (e.g. he has authored many papers with relatively few citations) or "influential" (e.g. most of his papers have an impact because they have received a significant number of citations).

The sequel is organized as follows. 
In the next section we will define two new specific areas in the citation curve. 
Based on these two new areas, we will establish two new metrics for evaluating the performance of authors in terms of impact. 
In Section \ref{sec:experiments} we will present our datasets, which were built by extracting data from the Microsoft Academic Search database. 
We will analyze these data to view the dataset characteristics.
Further, we will present the distributions of our new metrics for the above datasets as well as we will compare them with other metrics proposed in the literature.
Finally, in Section \ref{sec:results} we will present some of the resulting ranking tables based on the new metrics and \hi.
Section \ref{sec:conclusion} will conclude the article.

\section{Penalty Areas and New Indices}
\label{sec:areas}

Before proceeding further, in Table~\ref{symbols_tab} we summarize in a unifying way some symbols well-known from the literature, which will be used in the sequel.

\begin{table}[hbt]
\begin{tabular}{|c||l|}\hline
Symbol& Description \\ \hline \hline
$h$   & \hi of an author \\
$p$   & number of publications of an author \\
$P$   & set of publications of an author \\
$P_H$ & set of publications of an author that belong in the Core area \\
$P_T$ & set of publications of an author that belong in the Tail area \\
$p_T$ & number of publications that belong in $P_T$ \\
$C$   & number of citations of an author \\
$C_i$ & number of citations for publication $i$ \\
$C_H$ & number of citations for publications in $P_H$ \\
$C_T$ & number of citations for publications in $P_T$ \\
$C_E$ & number of citations in the upper area \\ \hline
\end{tabular}
\caption{Unified symbols and variables.}
\label{symbols_tab}
\end{table}

From the above symbols, it is apparent that the following expressions hold:
\begin{equation}
|P|~=~p
\end{equation}
\begin{equation}
|P_H|~=~h
\end{equation}
\begin{equation}
|P_T|~=~p_T~=~p-h
\end{equation}
\begin{equation}
C_H~=~\sum_{\forall i \in P_H}{C_i}
\end{equation}
\begin{equation}
C_T~=~\sum_{\forall i \in P_T}{C_i}
\end{equation}
\begin{equation}
C_E~=~C_H-h^2
\end{equation}

\subsection{The Tail Complement Penalty Area}
\label{sec:xarea}

\begin{figure}[bth]
\centering
\subfigure[]{\label{example_authorA}
\resizebox{5.5cm}{4.0cm}{\psfig{figure=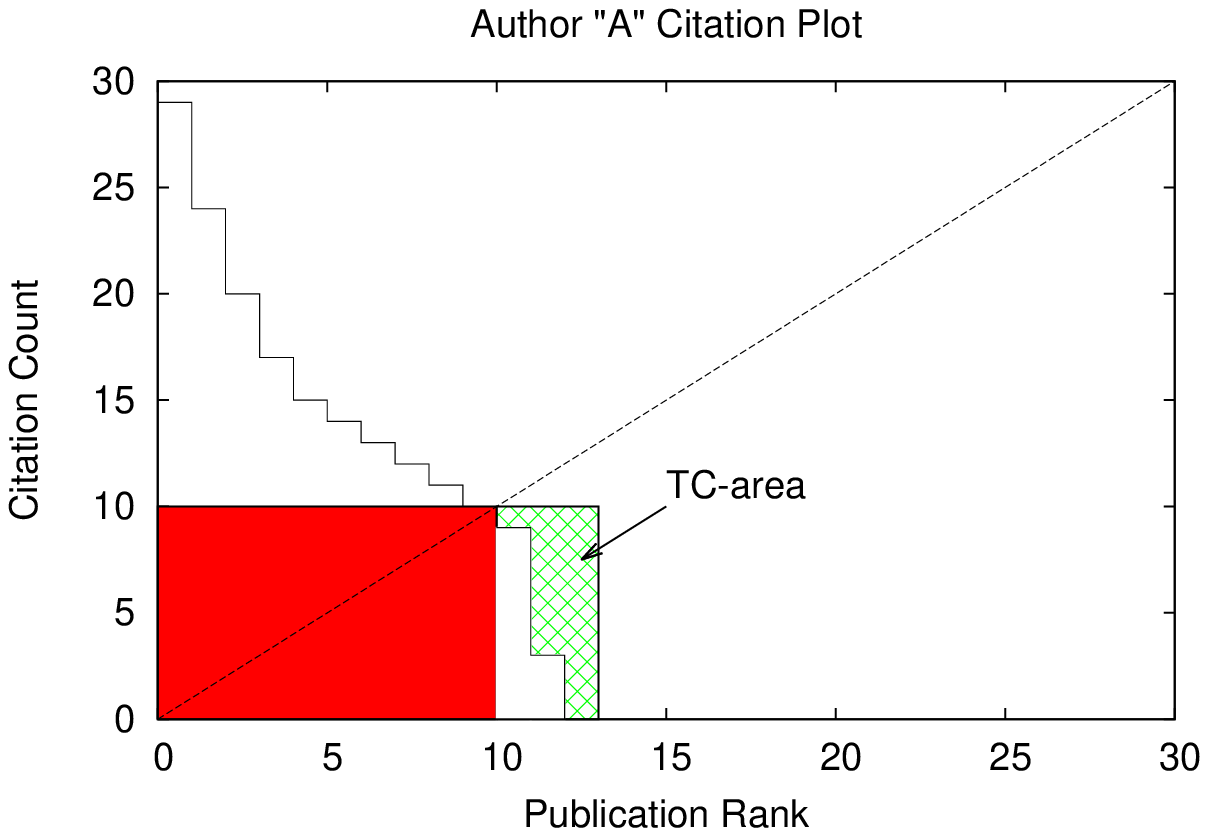,clip=}}
}
\subfigure[]{\label{example_authorB}
\resizebox{5.5cm}{4.0cm}{\psfig{figure=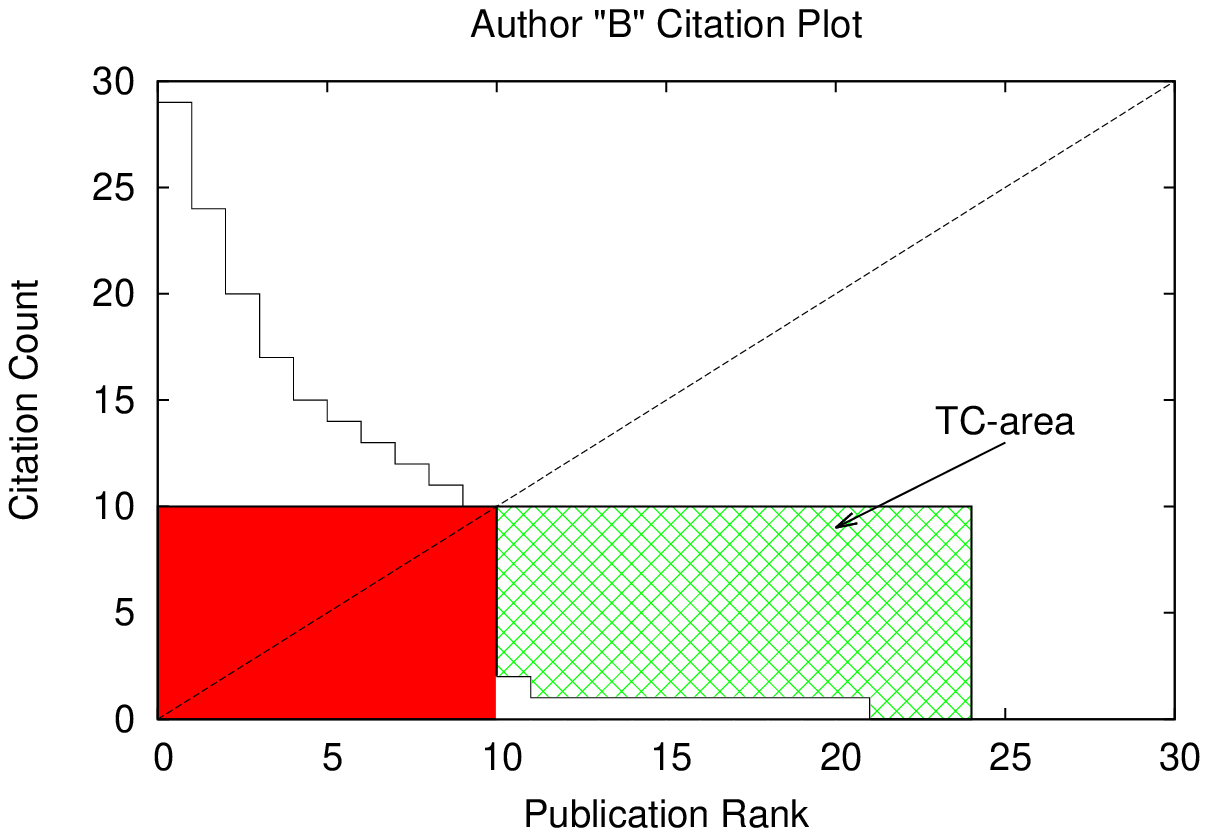,clip=}}
}
\caption{Citation curves of authors $A$ and $B$.}
\label{example_authorsAB}
\end{figure}

Let's consider the example of Figure~\ref{example_authorsAB}, which depicts the citation curves of two authors, $A$ and $B$. 
Both authors have the same macroscopic characteristics in terms of the number of citations, i.e. they have the same total number of citations $C$, identical core areas $C_H$ with \hi equal to 10, identical upper areas $C_E=65$, and the same number of citations in the tail area ($C_T=12$).

However, author $A$ has authored $p=13$ publications, whereas author $B$ has authored $p=24$ publications.
Also, for author $A$ it holds that $p_T=p-h=3$ and $C_T=12$ (the number of citations for his publications in the tail area are $\{9,3,0\}$ and in core area $\{29,24,20,17,15,14,13,12,11,10\}$).
On the other hand, for author $B$ it holds that $p_T=p-h=14$ and $C_T=12$ (the numbers of citations for his publications in the core area are the same as author $A$ and in the tail area are $\{2,1,1,1,1,1,1,1,1,1,1,0,0,0\}$).
It is definitely clear that since author $B$ has a significant number of publications in the tail area, he can be characterized as a "mass producer", whereas most of the publications of author $A$ have contributed to the calculation of the \hi, and thus the work of author $A$ has greater influence.

From the above example we understand that long tails reduce an author's influence.
Thus, we reach the conclusion that a long tail area should be considered as a negative characteristic when assessing a researcher's performance. 
For this purpose we define a new area, the {\it tail complement penalty area}, denoted as TC-area with size $C_{TC}$.
As shown in Figure~\ref{example_authorsAB}, the TC-area is much bigger for author $B$ than for author $A$.
More formally, we define the size of the tail complement penalty area as:
\begin{equation}
C_{TC} ~=~ \sum_{\forall i \in P_T}(h-C_i) ~=~ h \times (p-h) - C_T
\label{eq_C_X}
\end{equation}

\subsection{The Ideal Complement Penalty Area}
\label{sec:idealarea}

\begin{figure}[!tb]
\centerline{\psfig{figure=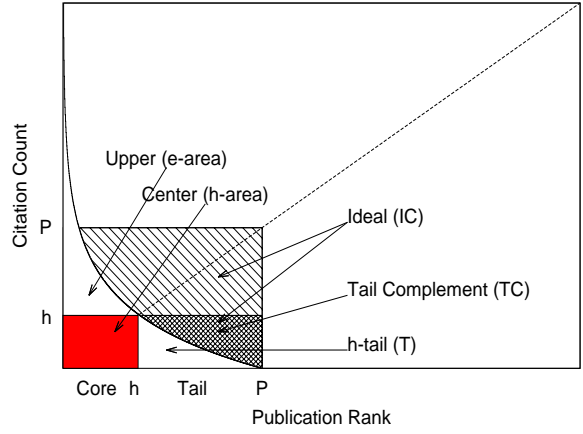,clip=,width=8cm,height=6cm}}
\caption{Citation curve with all areas defined.}
\label{example_dp_fig}
\end{figure}

"Ideally" an author could publish $p$ papers with $p$ citations each and get an \hi equal to $p$.
Thus, a square $p \times p$ could represent the minimum number of citations to achieve an \hi value equal to $p$.
Along the spirit of penalizing long tails, we can define another area in the citation curve: the {\it ideal complement penalty area} (IC-area), which is the complement of the citation curve with respect to the square $p \times p$.
Apparently, this area does not depend on \hi value as it happens for the case of TC-area.
Figure~\ref{example_dp_fig} shows graphically the IC-area. 
Notice that the IC-area includes the TC-area defined in the previous paragraph.
The size of the IC-area, $C_{IC}$, can be computed as follows: 
\begin{equation}
C_{IC} ~=~ \sum_{\forall i \in P, ~~C_i<p}{(p-C_i)}
\label{eq_CI}
\end{equation}

\subsection{New Metrics: $PT$ and $PI$}
\label{sec:PT}

To differentiate between authors with identical number of citations but different citation curves, we will introduce two new metrics. 
First, let us introduce the concept of {\it Parameterized Count}, $PC$, as follows: 
$$
PC ~=~ \kappa*h^2 + \varepsilon*C_E + \tau*C_T 
$$
where $\kappa,\varepsilon,\tau $ are integer values. Apparently:
\begin{itemize}
\item
if $\kappa=\varepsilon=\tau=1$, then it holds that $PC=C$, 
\item
if $\kappa$=1, $~\varepsilon,\tau=0$, then $PC=h^2$, 
\item
if $\varepsilon$=1, $~\kappa,\tau=0$, then $PC=C_E=e^2=C_h-h^2$, 
\item
if $\tau$=1, $~\kappa,\varepsilon=0$, then $PC=C_T$.
\end{itemize}
By assigning positive values to $\kappa,\varepsilon$ but negative values to $\tau$, we could favor authors with small tails in the citation curve. 
However, even this way we could not differentiate between the authors $A$ and $B$ of the previous example.

For this reason, instead of using the tail of the citation curve, we could use the tail complement penalty area. 
Thus, similarly to the previous equation we define the concept of {\it Penalty Index based on TC-area} as:
\begin{equation}
PT ~=~ \kappa*h^2 + \varepsilon*C_E - \sigma*C_{TC} 
\label{eq_PT}
\end{equation}
In the experiments that will be reported in the next sections, we will use the values of $\kappa=\varepsilon=\tau=\sigma=1$. 
Noticeably, it will appear that $PT$ can get negative values. 
Thus
\begin{itemize}
\item 
if an author has $PT<0$ then we can characterize him as {\it mass producer}. 
\item 
if an author has $PT>0$ then we can characterize him as {\it influential}
\end{itemize}

%The $\sqrt{PC_{X}}$ is comparable to \hi. 

In the same way as the $PT$, by taking into account the ideal complement penalty area we can define yet another penalizing metric as:
\begin{equation}
PI ~=~ \kappa*h^2 + \varepsilon*C_E + \tau*C_T - \iota*C_{IC}
\label{eq_PI}
\end{equation}
As well as the previous case, we will assume that $\kappa=\varepsilon=\tau=\iota=1$.
It will be mentioned in the experimental section that very few authors can have positive values for this metric.

By using the previously defined Penalty Indices the resulting values for authors $A$ and $B$ are shown in Table \ref{tab_exampleAB_authors_results}. 
Author $A$ has greater values than author $B$ for both $PI$ and $PT$ Penalty Indices. 
This is a desired result.

\begin{table}[!hbt]
\begin{center}
\begin{tabular}{|c||cccccccccc|} \hline
Author & $p$& $C$ & $h$& $C_T$& $C_E$& $C_H$& $C_{TC}$& $PT$& $C_{IC}$& $PI$ \\ \hline\hline
$A$    & 13 & 177 & 10 & 12   & 65   & 165  & 18      & 147 & 33      & 144 \\
$B$    & 24 & 177 & 10 & 12   & 65   & 165  & 128     & 37  & 404     & -227\\ \hline
\end{tabular}
\caption{Computed values for authors $A$ and $B$.}
\label{tab_exampleAB_authors_results}
\end{center}
\end{table}

\begin{figure}[!tb]
\centering
\subfigure[]{\label{h10_example4} \resizebox{8.25cm}{5.7cm}{\psfig{figure=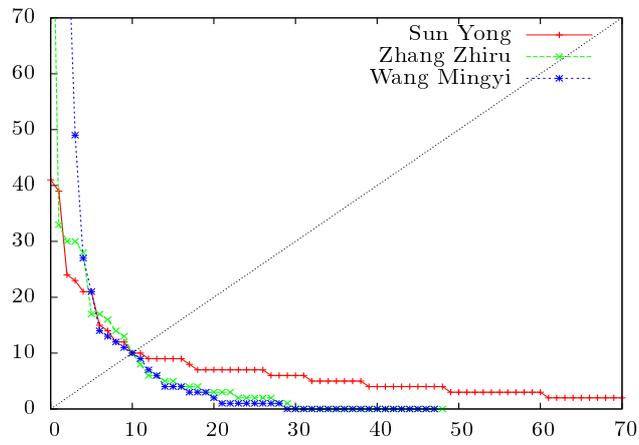,clip=}}}
%% 
%\subfigure[]{\label{h10-15_example5} \resizebox{5.5cm}{3.8cm}{ \psfig{figure=plots/real_examples/h10-15_example5.eps,clip=}}}
%% 
%\subfigure[]{\label{h10-15_example6} \resizebox{5.5cm}{3.8cm}{ \psfig{figure=plots/real_examples/h10-15_example6.eps,clip=}}}
%% 
\subfigure[]{\label{h10-15_example7} \resizebox{8.25cm}{5.7cm}{\psfig{figure=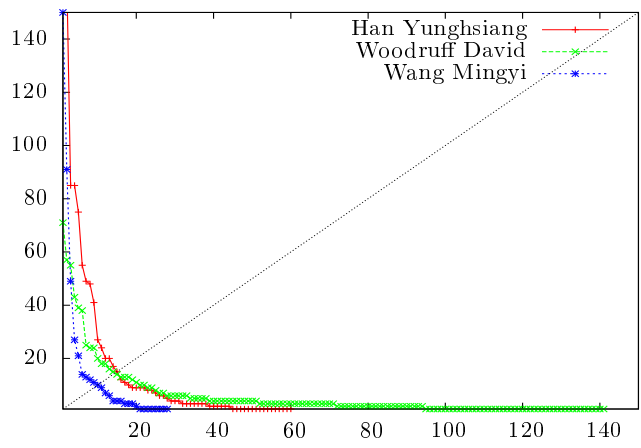,clip=}}}
\subfigure[The fig. \subref{h10-15_example7} adjusted.]{\label{h10-15_example7_scaled}
\resizebox{8.25cm}{5.7cm}{\psfig{figure=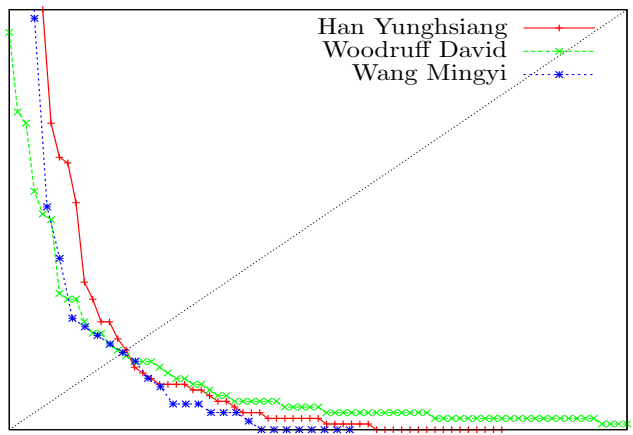,clip=}}}
\caption{Real examples.}
\label{real_examples_1}
\end{figure}

A further example with real data demonstrates the power of the new indices. 
In Table \ref{tab_real_examples} we present the raw data (i.e. \hi, number of publications $p$ and number of citations $C$) of 5 authors\footnote{We selected authors with relatively small number of publications and citations for better readability of the figures.} selected from Microsoft Academic Search\footnote{\url{http://academic.research.microsoft.com/}}. 
The last column shows the calculated $PT$ values, which can be positive as well as negative numbers. 
In Figure \ref{real_examples_1} we present some citation plots for these five authors.

\begin{table}[!bt]
\begin{center}

\begin{tabular}{|l@{}||r|r|r|r|}\hline
\multirow{2}{*}{\bf Author } & \multicolumn{1}{|c|}{\multirow{2}{*}{\bf $h$}}& \multicolumn{1}{|c|}{\multirow{2}{*}{\bf $p$}}& \multicolumn{1}{|c|}{\multirow{2}{*}{\bf $C$}}& \multicolumn{1}{|c|}{\multirow{2}{*}{\bf $PT$}}\\
 &  &  &  & \\\hline\hline
Han Yunghsiang		& 15 	& 105 		& 2040 		& 690 	\\ 
Woodruff David		& 15 	& 259 		& 1137 		& -2523 	\\ 
Wang Mingyi		& 10 	& 48 		& 1097 		& 717 	\\ 
Sun Yong		& 10 	& 319 		& 585 		& -2505 	\\ 
Zhang Zhiru		& 10 	& 49 		& 391 		& 1 	\\ 

\hline
\end{tabular}
\normalsize
\caption{Computed values for 5 authors.}
\label{tab_real_examples}
\end{center}
\end{table}

%\begin{table}[!hbt]
%\begin{center}
%\begin{tabular}{|l||cccc|} \hline
%Author         & \hi& $p$& $C$ & $PT$ \\ \hline\hline
%Wang Mingyi    & 10 & 48 & 1097& 717  \\ 
%Zhang Zhiru    & 10 & 49 & 391 & 1    \\ 
%Sun Yong       & 10 & 319& 585 & -2505\\ 
%Han Yunghsiang & 15 & 105& 2040& 690  \\ 
%Woodruff David & 15 & 259& 1137& -2523\\ \hline
%\end{tabular}
%\caption{Computed values for 5 authors.}
%\label{tab_real_examples}
%\end{center}
%\end{table}

%Also, in this table is reported the rank position for each author, the dataset is described 
%in section \ref{subsec:dataset}.
% In Figure \ref{h10_example4} are compared the authors: Sun Yong, Zhang Zhiru and Wang Ming-Yi. 
% 17987592: AuthorC: Mingyi     | Wang Wang Mingyi
% 3357373:  AuthorE: David      | Woodruff Woodruff David
% 1717316:  AuthorD: Yunghsiang | Han Han Yunghsiang 
% 12651485: AuthorA: Yong       | Sun Sun Yong
% 2312971:  AuthorB: Zhiru      | Zhang Zhang Zhiru

In Figure \ref{h10_example4} we compare three authors: Sun Yong, Zhang Zhiru and Wang Mingyi.
%In Figure \ref{h10_example4} are compared three authors: AuthorA, AuthorB and AuthorC.
%% They correspond to real names but we have preferred to present them anonymously.
They all have an \hi equal to 10. Sun Yong has the bigger and longest tail (red line). 
He actually has 319 publications but the citation curve is cropped to focus in the lower values. 
Apparently, he could be characterized as "mass producer". 
Actually, according to Table \ref{tab_real_examples} his $PT$ value equals -2505.
Zhang Zhiru (green line) has shorter tail than Sun Yong and higher excess area. 
From the same table we remark that his $PT$ value is 11 (i.e. close to zero). 
Finally, the last author of the example, Wang Mingyi (blue line), has similar tail with Zhang Zhiru but has bigger excess area ($e^2$). 
Definitely, he demonstrates the best citation curve out of the three authors of the example. 
In fact, his $PT$ score is 717, higher than the respective figure of the other two authors.

In Figure \ref{h10-15_example7}, again we compare three authors: Han Yunghsiang, Woodruff David and Wang Mingyi. 
The first two have \hi value equal to 15. 
Comparing the first two, it seems that Han Yunghsiang (red line) has better citation curve than Woodruff David (green line) because he has shorter tail and bigger excess area.
As a result the first one has $PT=717$, whereas the second one has $PT=-2523$.
Wang Mingyi (blue line) has smaller tail as well as quite big excess area but since there is a difference in \hi we cannot say for sure if he must be ranked higher or not than the others.

In Figure \ref{h10-15_example7_scaled} we have scaled the citation plots so that all lines do cut the line $y=x$ at the same point.
From this plot it is shown that Wang Mingyi has better curve than Woodruff David because he has shorter tail and bigger excess area.
When comparing Wang Mingyi to Han Yunghsiang, we see that the second one has longer tail but he also has bigger excess area.
Both curves show almost the same symmetry around the line $y=x$.
That is why they both have similar $PT$ values. 
This is a further positive outcome as authors with different quantitative characteristics (say, a senior and a junior one) may have similar qualitative characteristics, and thus classified together.

\section{Experiments}
\label{sec:experiments}

\subsection{DataSet}
\label{subsec:dataset}

During the period December 2012 to April 2013, we have produced 3 datasets. 
The first one consists of randomly selected authors (named "Random" thereof).
The second one includes highly productive authors (named "Productive").
The last one consists of authors in the top \hi list (named "Top h").
The publication and the citation data were extracted from the Microsoft Academic Search (MAS) database by using the MAS API\footnote{We appreciate the offer of Microsoft to gratis provide their database API.}. 

The dataset "Random" was generated as follows: 
We fetched a list of about 100000 authors belonging to the "Computer Science" Domain as reported by MAS. 
Note here that at least three sub-domains are assigned by MAS to every author. 
These three sub-domains may not belong all to the same domain (i.e. Computer Science). 
For example, an author may have two sub-domains from Computer Science and one from Medicine. 
We kept only the authors that have their first three sub-domains belonging to the domain of Computer Science. 
From this set we randomly selected 500 authors with at least 10 publications and at least 1 citation.

%We randomly selected 1000 authors from the DBLP database with more than 10 publications. 
%Then we identified the above authors in the MAS Database. 
%The number of authors that were identified was 953. 
%For these authors we downloaded the list of their publications and the appropriate meta-data.

The dataset "Productive" was generated with a similar way as previously. 
From the set of 100000 Computer Science authors we selected the top-500 most productive.
The less productive author from this sample has 354 publications.
%We randomly selected 500 authors from the DBLP with more than 150 publications. 
%436 authors were identified in the MAS Database. 

The third dataset named "Top h" was generated by querying the MAS Database for the top-500 authors in "Computer Science" domain ordered by \hi.

%The last one name "Mortals" was generated by fetching the list of "Computer Science" authors from MAS Database ordered by \hi and we randomly selected 400 with reported \hi between 15 and 25.

Table~\ref{table_stats} summarizes the information about our datasets with respect to the number of authors (line: \# of authors), number of publications (line: \# of publications), the number of citations (line: \# of Citations) and average/min/max numbers of citations and publications per author.

\begin{table}[!hbt]
\begin{center}
\begin{tabular}{|r|ccc|}\hline
 & \bf{Random} & \bf{Productive} & \bf{Top h}\\ \hline\hline
\# of authors     & 500   & 500    & 500 \\
\# of publications& 25679 & 223232 & 149462 \\
\#P/Author        & 51    & 446    & 298 \\
Min \#P/Author    & 10    & 354    & 92 \\
Max \#P/Author    & 768   & 1172   & 1172 \\
\# of Citations   & 410280& 3197880& 5015971 \\
\# Cit/Author     & 820   & 6395   & 10031 \\
Min \#Cit/Author  & 1     & 25     & 4405 \\
Max \#Cit/Author  & 47263 & 47263  & 47263\\ \hline
\end{tabular}
\caption{Statistics of the 3 samples}
\label{table_stats}
\end{center}
\end{table}

%In our set of experiments\footnote{Not all experiments reported in this article.} it is very important to know the year of publication for each article.
%Unfortunately about 92\% of the samples did have the meta data of publication year.
%In this point there was an extra effort to assume for each article the year of publication. 
%The algorithm is as follows:
%For each article A with no year meta data: Fetch all its references and all its citations (if they do exist). 
%If we found both references and citations for article A we assume that the publication year $Y$ should be grater or equal with the publication year of the newest article in Reference list ($max(RY)$) and less or equal with the publication year of the oldest article in Citation list ($min(CY)$).
%If $max(RY)==min(CY)$ we assume that $Y=max(RY)=min(CY)$.
%In case that the Reference list is empty we still may assume the publication year if the article receives at least 1 citation per year and it has at least 5 citations. In this case we assume that $Y=min(CY)$.
%By following the above algorithm we estimated the publication year for about 1.5-2\% of the publications in sample or about 15\% of the missed information. 
%The numbers are shown in Table~\ref{tab_stats}.
%
%For our experiments we ignore the publications with unknown and un-estimated year of publication. 
%Thus the values we compute about number of publications and \hi for each author do differ from those reported by MAS. 
%The value we compute are obviously smaller than the ones computed by MAS.

\subsection{Dataset Description}
\label{subsec:datadesc}

\begin{figure}[!htb]
\centering
%% 1
\subfigure[$h$-index]{\label{y0_c0_s1_distr_hindex_} \resizebox{5.5cm}{3.8cm}{ \psfig{figure=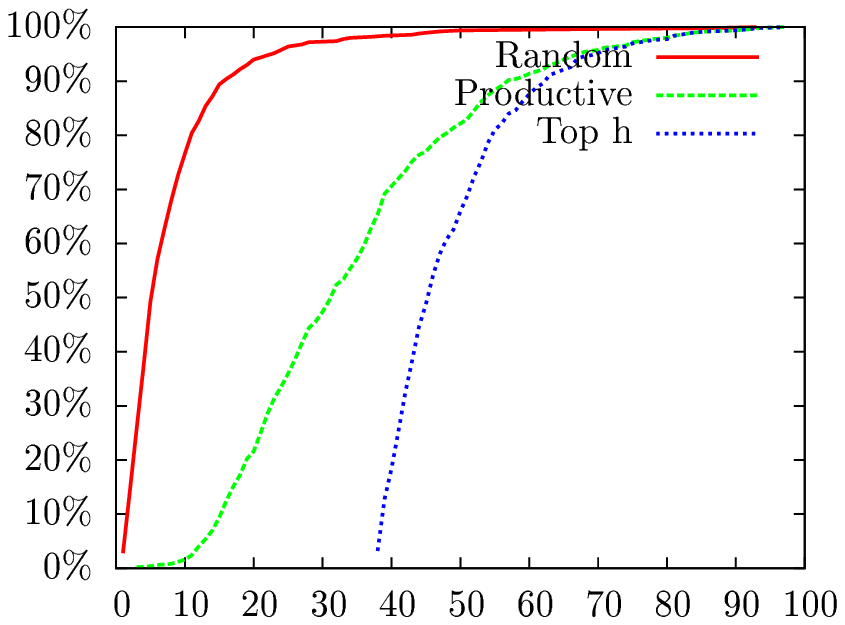,clip=}}}
\subfigure[$h$-index]{\label{y0_c0_s1_distr_hindex__grouped} \resizebox{5.5cm}{3.8cm}{ \psfig{figure=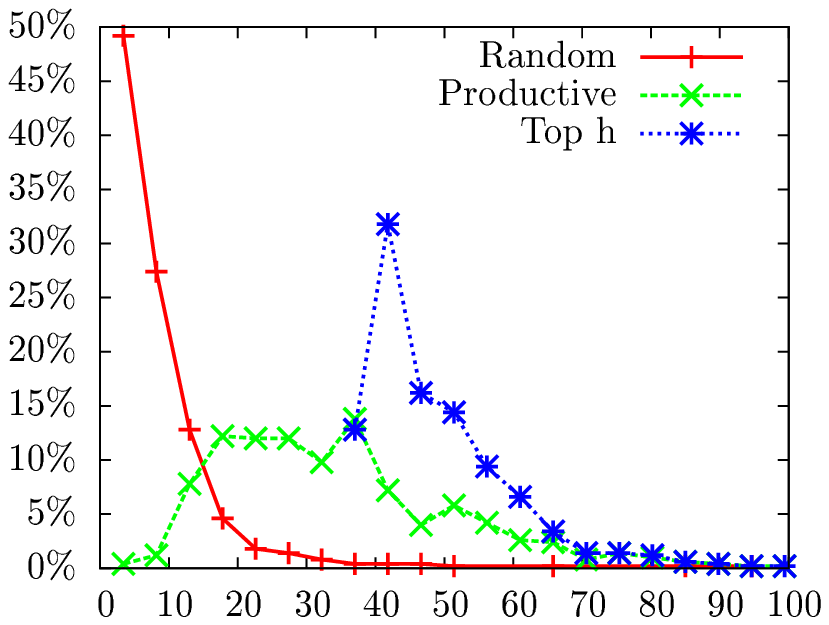,clip=}}}
%%% 2
%\subfigure[$a$]{\label{y0_c0_s1_distr_a_} \resizebox{5.5cm}{3.8cm}{ \psfig{figure=plots/density_all/y0_c0_s1_distr_a_.eps,clip=}}}
%\subfigure[$a$]{\label{y0_c0_s1_distr_a__grouped} \resizebox{5.5cm}{3.8cm}{ \psfig{figure=plots/density_all/y0_c0_s1_distr_a__grouped.eps,clip=}}}
%% 3
\subfigure[$C_S$ (* 10000)]{\label{y0_c0_s1_distr_C_S_} \resizebox{5.5cm}{3.8cm}{ \psfig{figure=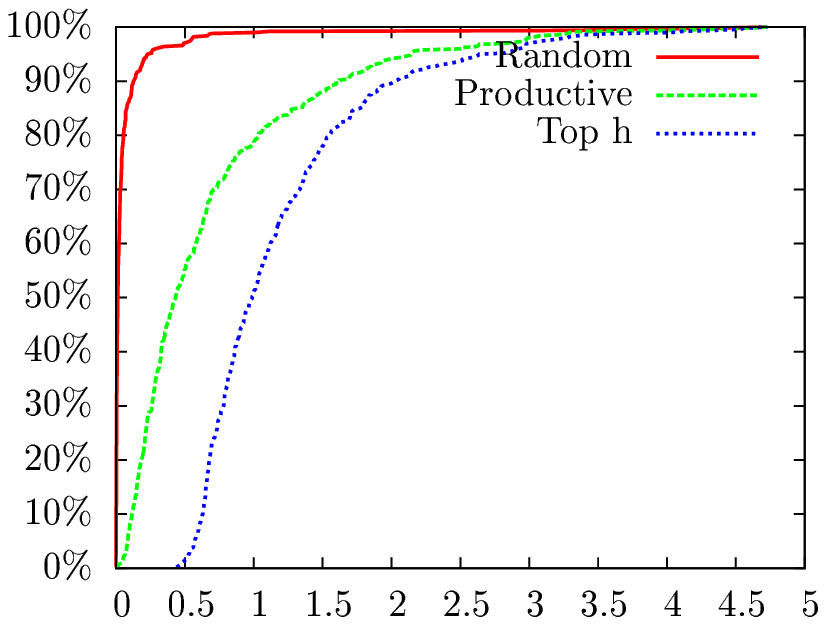,clip=}}}
\subfigure[$C_S$ (* 10000)]{\label{y0_c0_s1_distr_C_S__grouped} \resizebox{5.5cm}{3.8cm}{ \psfig{figure=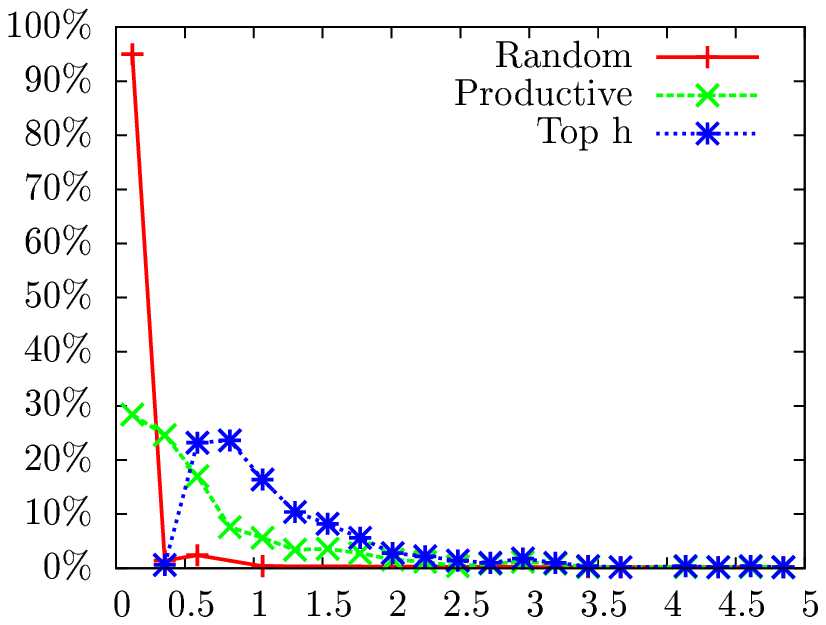,clip=}}}
%% 4
\subfigure[$m$]{\label{y0_c0_s1_distr_m_} \resizebox{5.5cm}{3.8cm}{ \psfig{figure=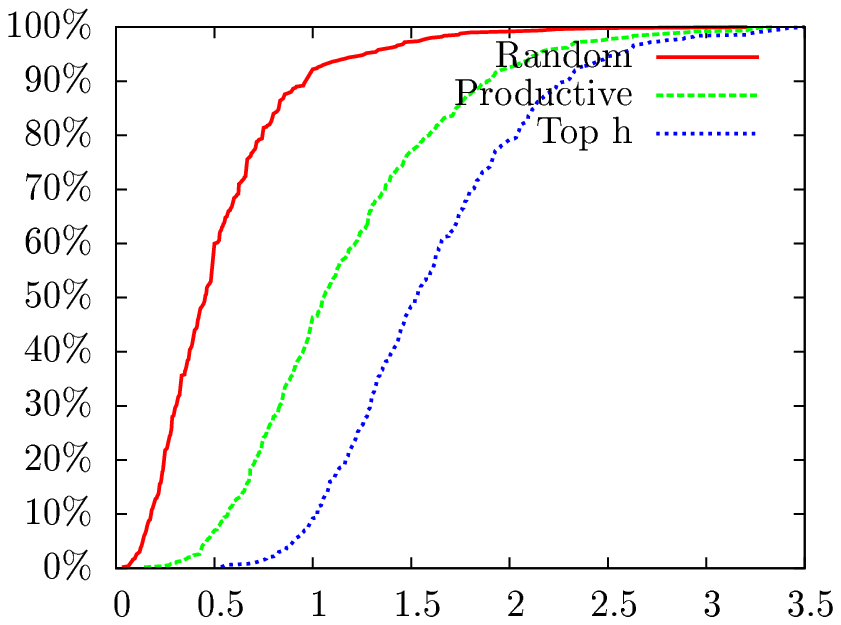,clip=}}}
\subfigure[$m$]{\label{y0_c0_s1_distr_m__grouped} \resizebox{5.5cm}{3.8cm}{ \psfig{figure=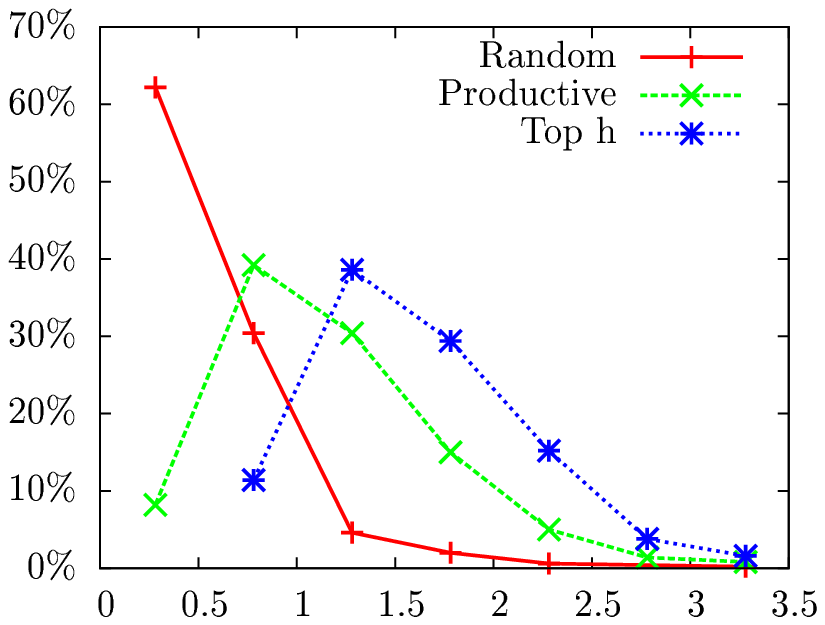,clip=}}}
\subfigure[$p$ (* 1000)]{\label{y0_c0_s1_distr_p_} \resizebox{5.5cm}{3.8cm}{ \psfig{figure=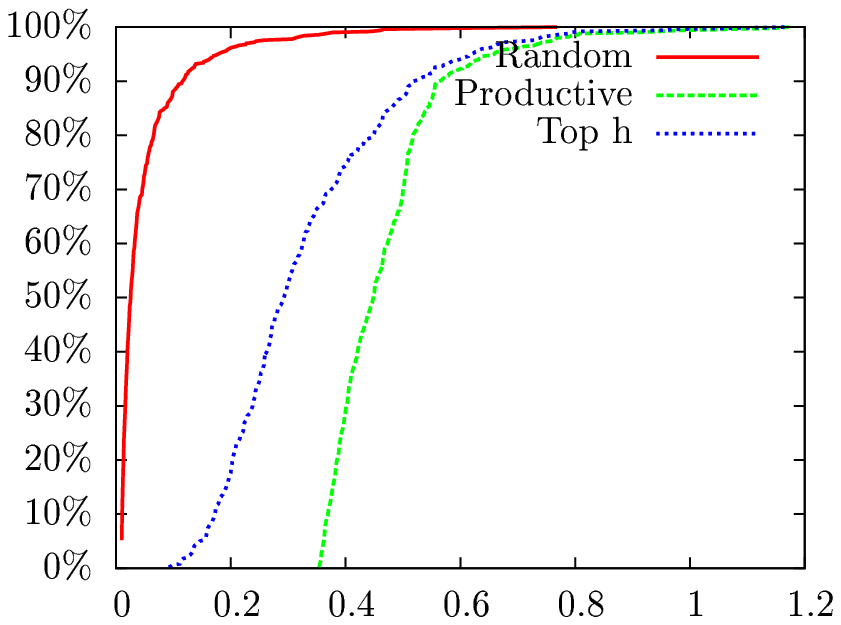,clip=}}}
\subfigure[$p$ (* 1000)]{\label{y0_c0_s1_distr_p__grouped} \resizebox{5.5cm}{3.8cm}{ \psfig{figure=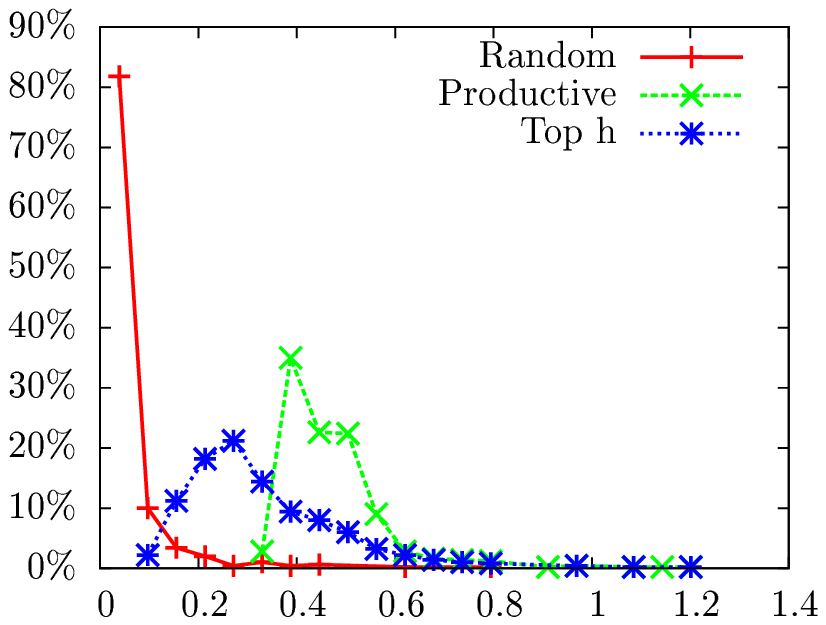,clip=}}}

\caption{Distributions of \hi, $m$, $p$, $C_S$ (left, cummulative)}
\label{fig_plot_all_plot_basic}
\end{figure}

Figure~\ref{fig_plot_all_plot_basic} shows the distributions for the values of \hi, $m$, $C$ and $p$. 
Plots are illustrated in pairs. The left ones show cumulative distributions.
For example, in Figure~\ref{y0_c0_s1_distr_hindex_} we see that 80\% of the authors in the sample "Random" (red line) have \hi less than 10. 
It is obvious that the sample "Top h" (blue line) has higher values for the \hi. 
Figures \ref{y0_c0_s1_distr_C_S_} and \ref{y0_c0_s1_distr_C_S__grouped} show the distributions for the total number of citations. 
As expected the sample "Top h" has the highest values.
Figures \ref{y0_c0_s1_distr_m_} and \ref{y0_c0_s1_distr_m__grouped} show the distributions for the $m$ value as defined by Hirch in \cite{hirsch05}:

\textit{A value of $m\approx 1$ (i.e., an \hi of 20 after 20 years of scientific activity), characterizes a successful scientist.
A value of $m \approx 2$ (i.e., an \hi of 40 after 20 years of scientific activity), characterizes outstanding scientists, likely to be found only at the top universities or major research laboratories.
A value of $m \approx 3$ or higher (i.e., an \hi of 60 after 20 years, or 90 after 30 years), characterizes truly unique individuals.}

The above statement is verified in these figures; only a few authors have $m>3$.
Figures \ref{y0_c0_s1_distr_p_} and \ref{y0_c0_s1_distr_p__grouped} illustrate the distributions for the total number of publications.
It is obvious that in the "Random" sample (red line) there are relatively low values for the total number of publications. 
Also, as expected the distribution for the "Productive" sample has the highest values for the total number of publications.

%On the other hand the distributions of the total number of publications for the second and third sample are almost equal. 
%By taking into account the difference in the \hi distribution, this shows that in the third sample ("Top h") there are scientists with definitely higher quality than in the second sample.

We have conducted further experiments to study the behavior of other common factors like $\alpha$ \cite{hirsch05} and $e^2$ \cite{Zhang09}. 
However, the results did not show to carry any additional noticeable information, and, thus, figures for these factors are not included here.

%The excess area (Figures \ref{y0_c0_s1_distr_e2_} and \ref{y0_c0_s1_distr_e2__grouped}) has similar distributions with the \hi ones for the third sample ("Top \hi Authors"). 
%For the first and second sample, the values are much lower than in the \hi. 
%This is almost obvious, since the "Top \hi Authors" is expected to have higher values for $e^2$.
%Figure~\ref{qqplots_hindex_c0_s1_vs_e2_c0_s1_all} shows a q-q plot comparing the ranking based on $e^2$ and \hi. 
%This plot also shows that the ranking do have great difference for the first two samples. 
%
%The distributions for $C_H$ (Figures~\ref{y0_c0_s1_distr_C_H_} and \ref{y0_c0_s1_distr_C_H__grouped}) are very similar to the $e^2$ ones. 
%This is also shown in Figure~\ref{qqplots_e2_c0_s1_vs_C_H_c0_s1_all}, where a comparative q-q plot for these two ranking is displayed.

\subsection{Do we need new Indices?}
\label{subsec:need}

\begin{figure}[!htb]
\centering
%\subfigure[$h$ vs. $C_S$]{\label{qqplots_hindex_c0_s1_y0_vs_C_S_c0_s1_y0_all} \resizebox{5.5cm}{3.15cm}{ \psfig{figure=plots/qqplots/qqplots_hindex_c0_s1_y0_vs_C_S_c0_s1_y0_all.eps,clip=}}}
%% 
\subfigure[$h$ vs. $C_S$ (unioned)]{\label{qqplots_hindex_c0_s1_y0_vs_C_S_c0_s1_y0_union} \resizebox{5.5cm}{3.9cm}{ \psfig{figure=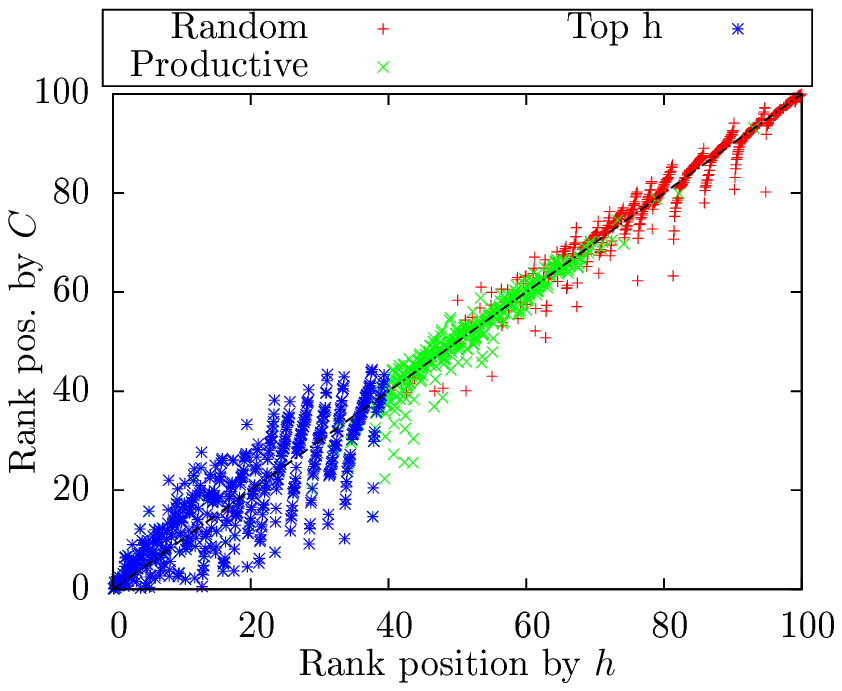,clip=}}}
%% 
%\subfigure[$h$ vs. $PT$]{\label{qqplots_hindex_c0_s1_y0_vs_PT_c0_s1_y0_all} \resizebox{5.5cm}{3.15cm}{ \psfig{figure=plots/qqplots/qqplots_hindex_c0_s1_y0_vs_PT_c0_s1_y0_all.eps,clip=}}}
%% 
\subfigure[$h$ vs. $PT$ (unioned)]{\label{qqplots_hindex_c0_s1_y0_vs_PT_c0_s1_y0_union} \resizebox{5.5cm}{3.9cm}{ \psfig{figure=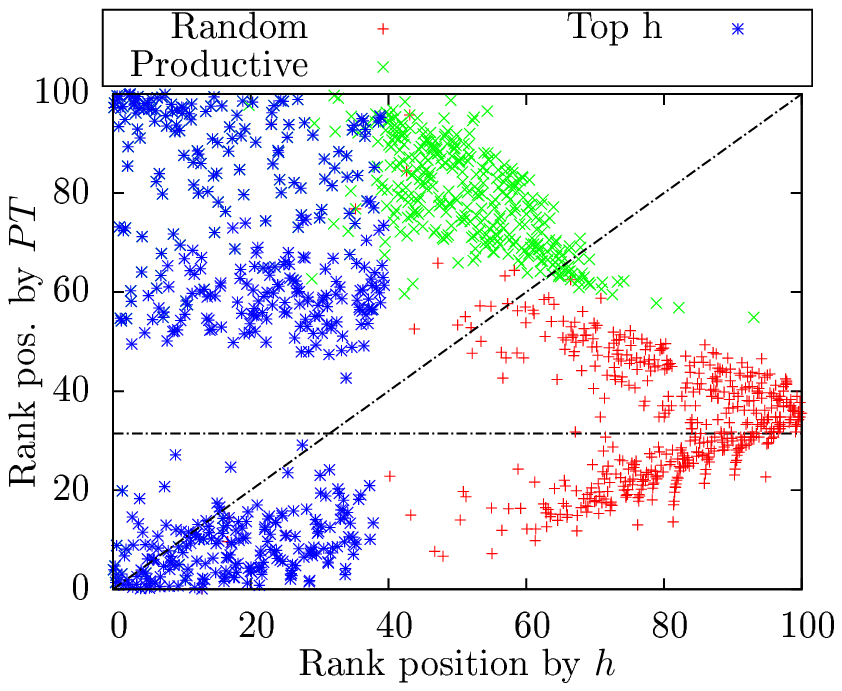,clip=}}}

%% 
%\subfigure[$e^2$ vs. $PT$]{\label{qqplots_e2_c0_s1_y0_vs_PT_c0_s1_y0_all} \resizebox{5.5cm}{3.15cm}{ \psfig{figure=plots/qqplots/qqplots_e2_c0_s1_y0_vs_PT_c0_s1_y0_all.eps,clip=}}}
%% 
% \subfigure[$e^2$ vs. $PT$ (unioned)]{\label{qqplots_e2_c0_s1_y0_vs_PT_c0_s1_y0_union} \resizebox{5.5cm}{3.9cm}{ \psfig{figure=plots/qqplots/qqplots_e2_c0_s1_y0_vs_PT_c0_s1_y0_union.eps,clip=}}}
%% 
%\subfigure[$C_S$ vs. $PT$]{\label{qqplots_C_S_c0_s1_y0_vs_PT_c0_s1_y0_all} \resizebox{5.5cm}{3.15cm}{ \psfig{figure=plots/qqplots/qqplots_C_S_c0_s1_y0_vs_PT_c0_s1_y0_all.eps,clip=}}}
%% 
\subfigure[$C_S$ vs. $PT$ (unioned)]{\label{qqplots_C_S_c0_s1_y0_vs_PT_c0_s1_y0_union} \resizebox{5.5cm}{3.9cm}{ \psfig{figure=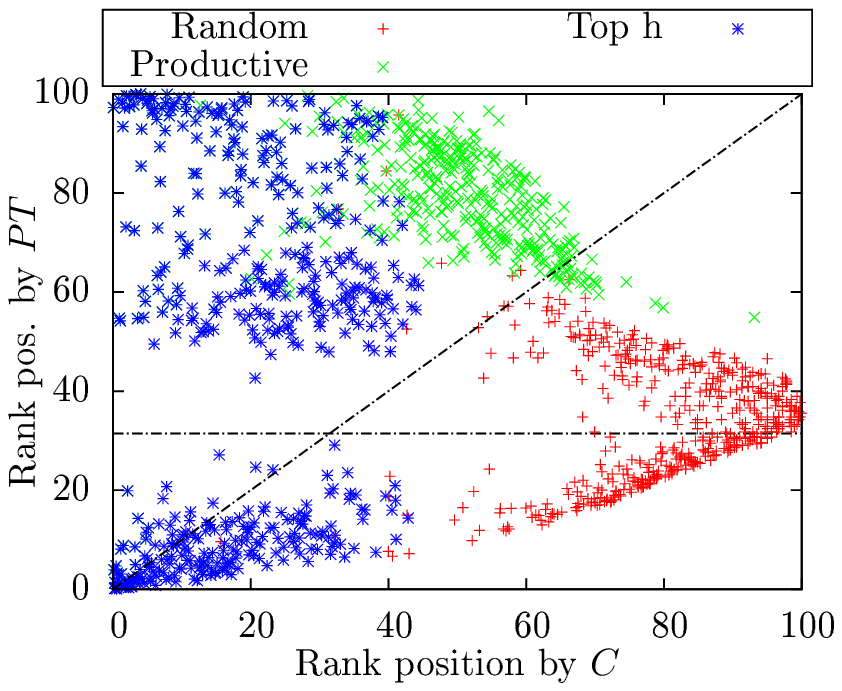,clip=}}}
%% 
%\subfigure[$h$ vs. $PI$]{\label{qqplots_hindex_c0_s1_y0_vs_PI_c0_s1_y0_all} \resizebox{5.5cm}{3.15cm}{ \psfig{figure=plots/qqplots/qqplots_hindex_c0_s1_y0_vs_PI_c0_s1_y0_all.eps,clip=}}}
% \subfigure[$h$ vs. $PI$ (unioned)]{\label{qqplots_hindex_c0_s1_y0_vs_PI_c0_s1_y0_union} \resizebox{5.5cm}{3.9cm}{ \psfig{figure=plots/qqplots/qqplots_hindex_c0_s1_y0_vs_PI_c0_s1_y0_union.eps,clip=}}}

\subfigure[$cit/p$ vs. $PT$ (unioned)]{\label{qqplots_citp_c0_s1_y0_vs_PT_c0_s1_y0_union} \resizebox{5.5cm}{3.8cm}{ \psfig{figure=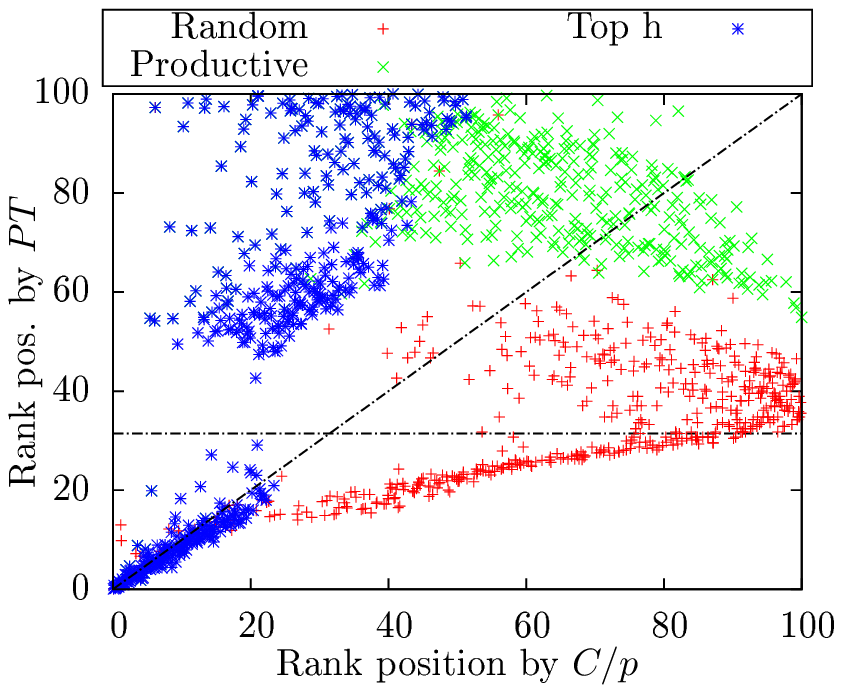,clip=}}}
% \subfigure[$cit/p$ vs. $PI$ (unioned)]{\label{qqplots_citp_c0_s1_y0_vs_PI_c0_s1_y0_union} \resizebox{5.5cm}{3.8cm}{ \psfig{figure=plots/qqplots/qqplots_citp_c0_s1_y0_vs_PI_c0_s1_y0_union.eps,clip=}}}
\subfigure[$cit/p$ vs. $h$ (unioned)]{\label{qqplots_citp_c0_s1_y0_vs_hindex_c0_s1_y0_union} \resizebox{5.5cm}{3.8cm}{ \psfig{figure=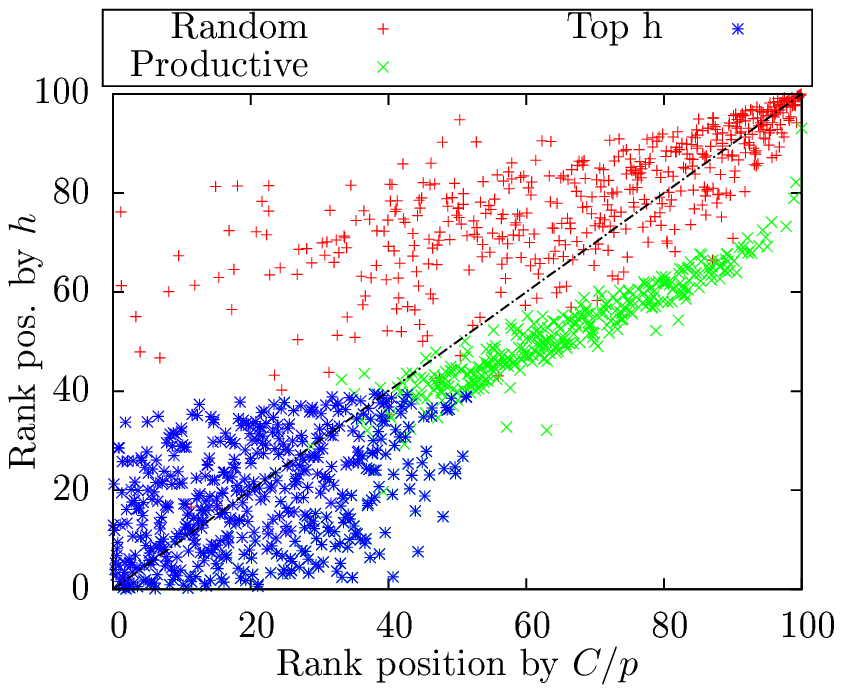,clip=}}}

\caption{Q-Q plots: X- and Y-axis denote normalized rank positions (\%)}
\label{plot_all_qqplot_1}
\end{figure}

In this section we perform some comparisons to show that our newly defined indices differ from existing ones. 
Actually our new metrics separates the rank tables into two parts independently from the rank positions.

% In Figure~\ref{plot_all_qqplot_1} are presented some q-q plots to compare different types of rankings and their behavior.

% In Fig~\ref{qqplots_hindex_c0_s1_y0_vs_C_S_c0_s1_y0_all} x-axis denotes the rank position (normalized in percent) of an author by \hi. 
% The y-axis denotes the rank position by the total number of citations ($C$). 
% A point at $(x,y)$ denotes that an author is ranked at the $x$-th place by the x-axis metric (in this figure \hi), and at $y$-th place by the y-axis metric.
% In Figure~\ref{qqplots_hindex_c0_s1_y0_vs_C_S_c0_s1_y0_all} all samples are ranked separately, so there are six rank tables; two rank tables for each sample. 
% The points for each sample are presented with different colors, i.e. a blue point at (30,50) means that an authors that belongs to sample "Top h" is ranked 30-th by \hi and 50-th by the total number of citations.

In Figure~\ref{qqplots_hindex_c0_s1_y0_vs_C_S_c0_s1_y0_union} 
%is of the same type of plot but we have only one rank table for all the samples. 
the x-axis denotes the rank position (normalized percentagewise) of an author by \hi, whereas the y-axis denotes the rank position by the total number of citations ($C$). 
%So, there are only two rank tables, one table ranked by the x-axis metric (\hi) and one table ranked by the y-axis metric (total number of citations).
Each point denotes the positions of an author ranked by the two metrics. 
Note, that all three samples are merged but if the point is blue, then the author belongs to the "Top h" sample, if the point is green then he belongs to "Productive" sample etc. 
If an author belongs to more than one sample, then only one color is visible since the bullet overwrites the previous one.
From Figure \ref{qqplots_hindex_c0_s1_y0_vs_C_S_c0_s1_y0_union} the outcomes are:
\begin{itemize}
\item 
"Top h" authors are ranked in the first 40\% of the rank table by \hi, as well as in the top 40\% by the total number of citations ($C$).
\item 
"Productive" authors are mainly ranked by \hi between 30\% and 70\%.
The rank positions by $C$ are between 20\% and 70\%.
%\item 
%"Mortals" are ranked between 27\% and 60\% by \hi and between 20\% and 80\% by $C$.
\item 
"Random" authors are mainly ranked below 60\% for both metrics with some outliers in the range 0-60\%, mostly by $C$.
\end{itemize}
All the above remarks may seem expected for both \hi and $C$ ranking.
Also, it comes out that the \hi ranking does not differ significantly from the $C$ ranking; i.e. they are correlated.

In Figure \ref{qqplots_hindex_c0_s1_y0_vs_PT_c0_s1_y0_union} the \hi ranking is compared to $PT$ ranking. 
It can be seen that there is no obvious correlation between $PT$ and \hi.
% Figure \ref{qqplots_hindex_c0_s1_y0_vs_PT_c0_s1_y0_union} gives a better view of what is happening when we produce only one rank table for all samples.
Note that the horizontal line at about 32\% (also, later shown in Table \ref{positive_tab}) shows the cut point of $PT$ for the zero value. 
Authors that reside below this line have $PT>0$ and authors above this line have $PT<0$.
\begin{itemize}
\item 
"Top h" authors are split to two groups. The first group is ranked in the top 20\% of the $PT$ rank table. 
The second group is ranked in the last 50\%. These two groups are also separated by the zero line of $PT$.
\item 
"Productive" authors are almost all ranked at lower positions by $PT$ than by \hi. 
Almost all points reside above the $PT$ zero line and also above the line $y=x$ (with some exceptions at about 65-70\% of the rank list).
%\item 
%The "Mortals" are ranked at higher positions by $PT$ rather than \hi but they are also split by the line $PT=0$.
\item 
"Random" authors are also generally higher ranked by $PT$ than by \hi. 
They are also split into two groups by the line $PT=0$. 
\end{itemize}
From all these statements, it seems that $PT$ is not correlated to \hi, whereas the line $PT=0$ plays the role of a symmetric axis. 
Thus, it emerges as the key value that separates the "Influential" authors from the "Mass producers".

Also, in Figure \ref{qqplots_C_S_c0_s1_y0_vs_PT_c0_s1_y0_union} we compare $PT$ ranking against $C$ (total number of citations) ranking. 
It is expected that the plot would be similar to Figure \ref{qqplots_hindex_c0_s1_y0_vs_PT_c0_s1_y0_union} based on the similarity of \hi with $C$.

In Figures \ref{qqplots_citp_c0_s1_y0_vs_PT_c0_s1_y0_union} and \ref{qqplots_citp_c0_s1_y0_vs_hindex_c0_s1_y0_union} we compare \hi and $PT$ with the average number of citations per publication ($cit/p$) ranking.
It is apparent that $PT$ is not correlated to $cit/p$. 
\hi is also uncorrelated to $cit/p$, however the points of the qq-plot in Figure \ref{qqplots_citp_c0_s1_y0_vs_hindex_c0_s1_y0_union} are closer to the line $x=y$ than the points of Figure \ref{qqplots_citp_c0_s1_y0_vs_PT_c0_s1_y0_union}.

Conclusively, the $PT$ ranking is not correlated to \hi, $C$ and $cit/p$. 
Similar graphs were produced from additional experimental comparisons that we have performed. 
Therefore, for brevity we do not include these graphs in this report. 

% We see the same exactly behavior when comparing $PT$ to $e^2 (C_E)$ in Figures \ref{qqplots_e2_c0_s1_y0_vs_PT_c0_s1_y0_all} and \ref{qqplots_e2_c0_s1_y0_vs_PT_c0_s1_y0_union}
% as well as to $C$ (number of citations) in Figures \ref{qqplots_C_S_c0_s1_y0_vs_PT_c0_s1_y0_all} and \ref{qqplots_C_S_c0_s1_y0_vs_PT_c0_s1_y0_union}. 
%We see the same exactly behavior when comparing $PT$ to $e^2 (C_E)$ in Figure \ref{qqplots_e2_c0_s1_y0_vs_PT_c0_s1_y0_union} as well as to $C$ (number of citations) in Figure \ref{qqplots_C_S_c0_s1_y0_vs_PT_c0_s1_y0_union}. 
%The line $PT=0$ splits the authors into two groups. 
%Note that there as some authors that are ranked higher by $PT$ than $e^2$ (or $C$) but they reside in the negative group. 
%Also there are authors that are ranked in lower position by $PT$ than $e^2$ (or $C$) but they reside in the positive group.
%Also in Figures \ref{qqplots_hindex_c0_s1_y0_vs_PI_c0_s1_y0_all} and \ref{qqplots_hindex_c0_s1_y0_vs_PI_c0_s1_y0_union} is presented the comparison between \hi and $PI$. 

% Also in Figure \ref{qqplots_hindex_c0_s1_y0_vs_PI_c0_s1_y0_union} is presented the comparison between \hi and $PI$. 
% The layout is similar to that \hi vs. $PT$ but the line $PI=0$ is much lower (at 7\%).

\subsection{Statistical Analysis}
\label{subsec:stats}

\begin{figure}[!hbt]
\centering
%% 8
\subfigure[$C_T$ (* 1000)]{\label{y0_c0_s1_distr_C_T_} \resizebox{5.5cm}{3.1cm}{ \psfig{figure=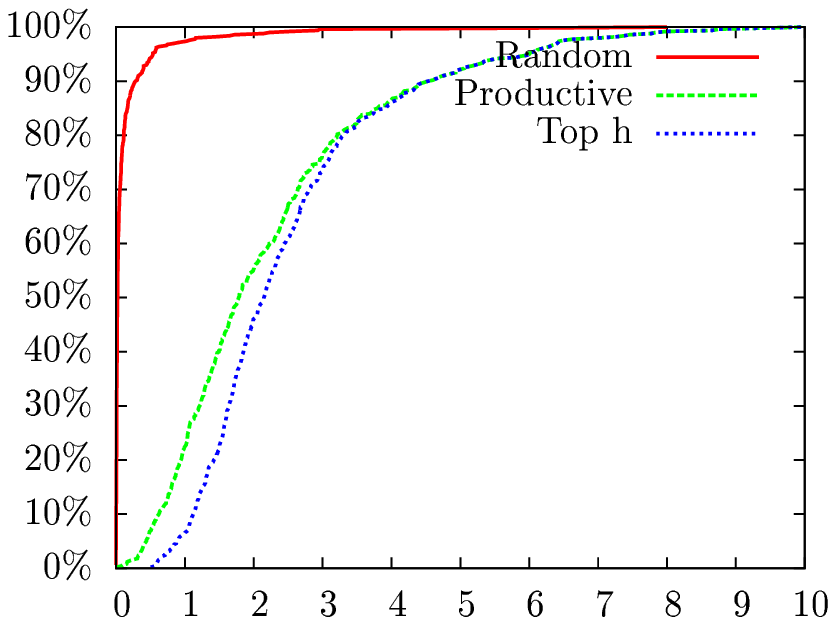,clip=}}}
\subfigure[$C_T$ (* 1000)]{\label{y0_c0_s1_distr_C_T__grouped} \resizebox{5.5cm}{3.1cm}{ \psfig{figure=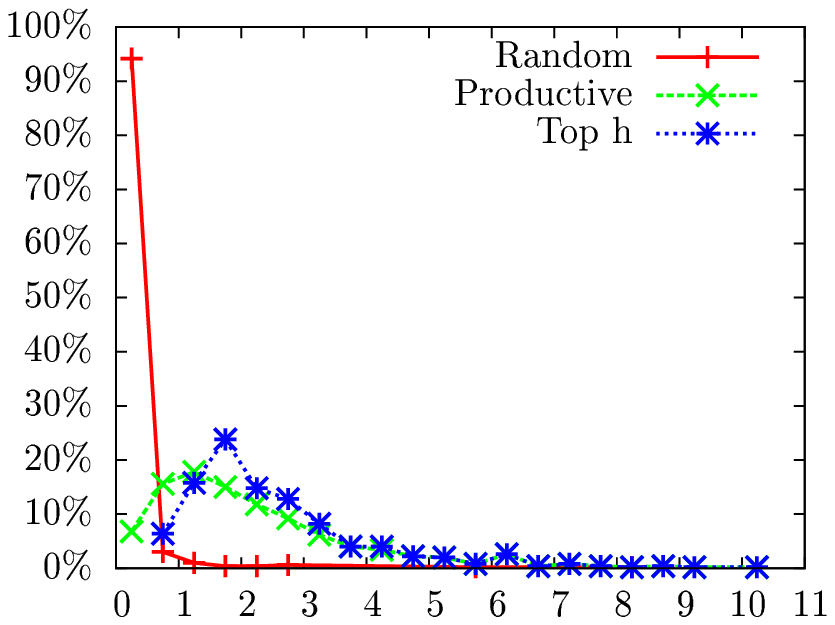,clip=}}}
%% 9
\subfigure[$C_{TC}$ (* 10000)]{\label{y0_c0_s1_distr_C_X_} \resizebox{5.5cm}{3.1cm}{ \psfig{figure=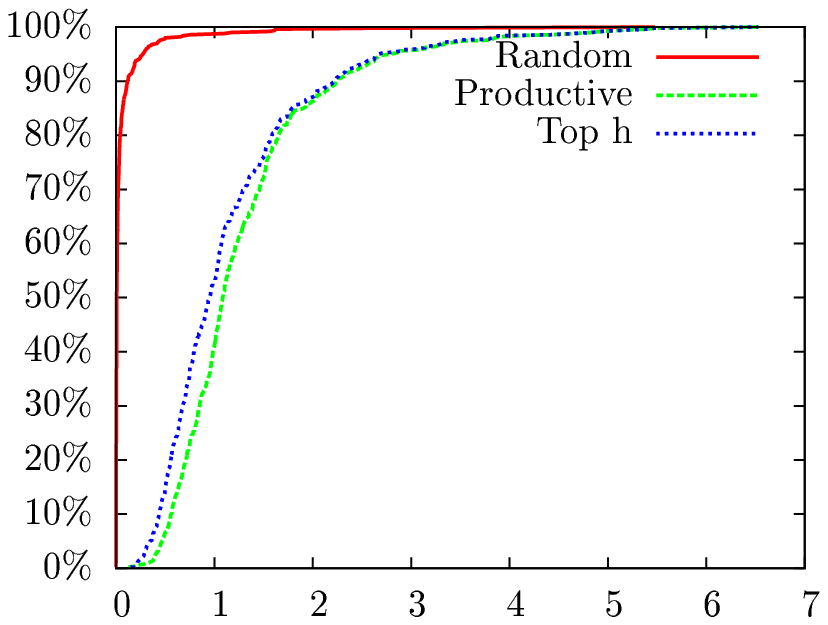,clip=}}}
\subfigure[$C_{TC}$ (* 10000)]{\label{y0_c0_s1_distr_C_X__grouped} \resizebox{5.5cm}{3.1cm}{ \psfig{figure=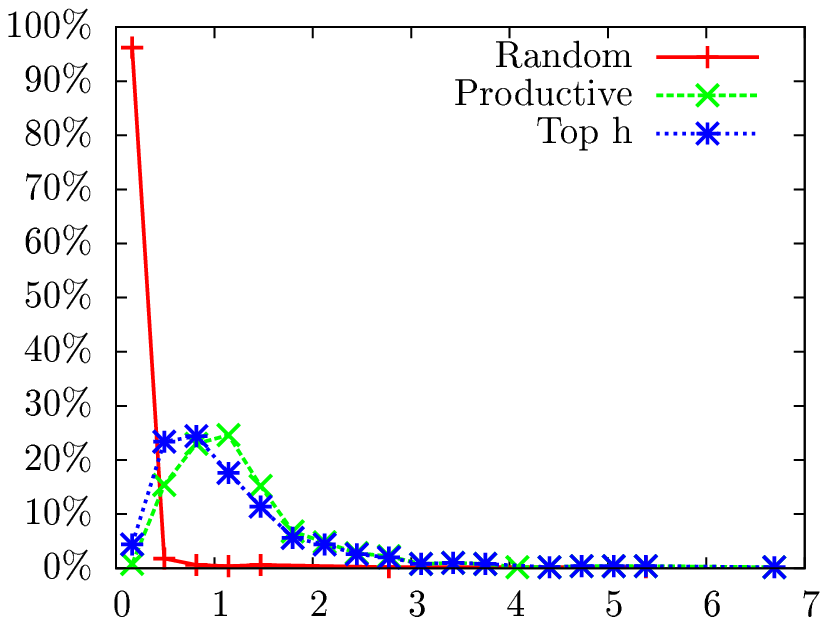,clip=}}}
%% 10
\subfigure[$C_{IC}$ (* 100000)]{\label{y0_c0_s1_distr_C_I_} \resizebox{5.5cm}{3.1cm}{ \psfig{figure=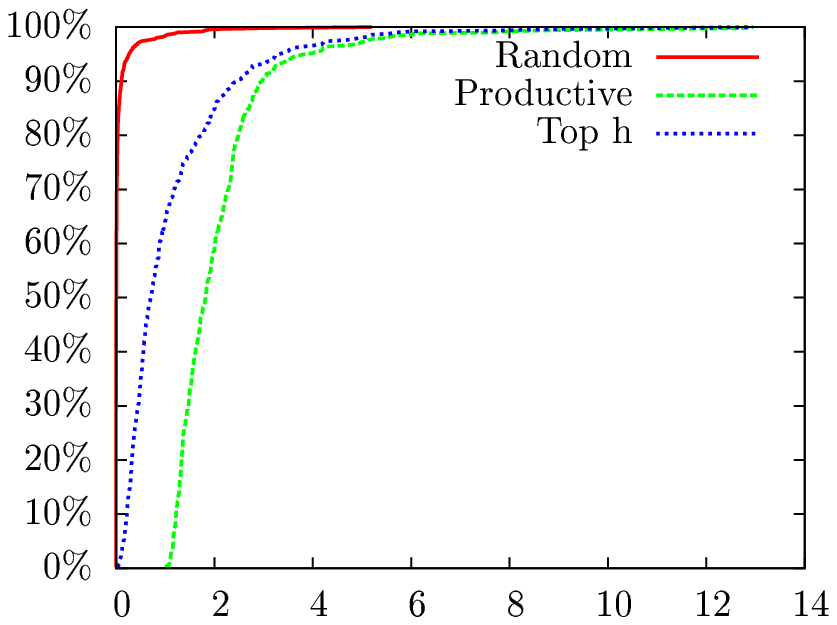,clip=}}}
\subfigure[$C_{IC}$ (* 100000)]{\label{y0_c0_s1_distr_C_I__grouped} \resizebox{5.5cm}{3.1cm}{ \psfig{figure=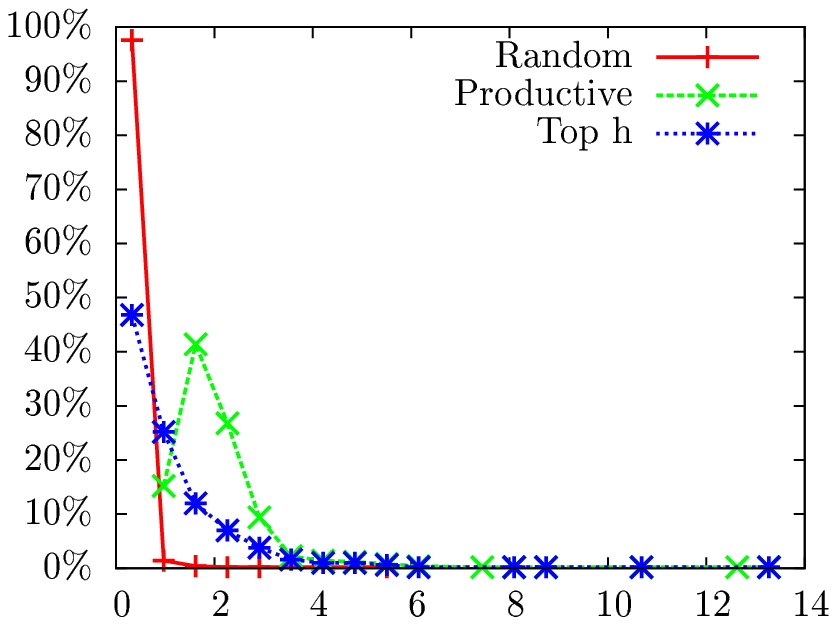,clip=}}}
%
%\label{fig_plot_all_plot_our_areas1}
%\caption{Distributions of $C_T$, $C_X$ (Expectation area), $C_{I}$ (Dist. to Perfection area)}
%\end{figure}
%\begin{figure}[!ht]
%
%% 11
% \subfigure[$C_{P}$ (* 10000)]{\label{y0_c0_s1_distr_C_P_} \resizebox{5.5cm}{3.1cm}{ \psfig{figure=plots/density_all/y0_c0_s1_distr_C_P_.eps,clip=}}}
% \subfigure[$C_{P}$ (* 10000)]{\label{y0_c0_s1_distr_C_P__grouped} \resizebox{5.5cm}{3.1cm}{ \psfig{figure=plots/density_all/y0_c0_s1_distr_C_P__grouped.eps,clip=}}}
%% 12
% \subfigure[$C_{M}$ (* 10000)]{\label{y0_c0_s1_distr_C_M_} \resizebox{5.5cm}{3.1cm}{ \psfig{figure=plots/density_all/y0_c0_s1_distr_C_M_.eps,clip=}}}
% \subfigure[$C_{M}$ (* 10000)]{\label{y0_c0_s1_distr_C_M__grouped} \resizebox{5.5cm}{3.1cm}{ \psfig{figure=plots/density_all/y0_c0_s1_distr_C_M__grouped.eps,clip=}}}
% \caption{Distributions of $C_T$, $C_X$ (Expectation area), $C_{I}$ (Dist. to Perfection area), $C_P$, $C_M$}
% Removed C_P and C_M for this version

\caption{Distributions of $C_T$, $C_{TC}$ (tail complement), $C_{IC}$ (ideal complement)}
\label{fig_plot_all_plot_our_areas2}
\end{figure}

Figure~\ref{fig_plot_all_plot_our_areas2} shows the distributions for the areas defined in the previous.
In particular, Figures~\ref{y0_c0_s1_distr_C_T_} and \ref{y0_c0_s1_distr_C_T__grouped} illustrate the distributions for the $C_T$ (tail) area. 
It seems that the "Top h" cumulative distribution is very similar to the "Productive" one, however, the "Top h" distribution has slightly higher values.
%The "Mortals" sample is very close to the "Random" one. 
%On the other hand it seems that "Top h" has higher values for $C_T$ than "Productive". 
%Having in mind that these two samples have similar distribution for the total number of publications, this means that authors in "Top h" have more citations in the tail area than authors in "Productive".

Figures~\ref{y0_c0_s1_distr_C_X_} and \ref{y0_c0_s1_distr_C_X__grouped} illustrate the distributions for the $C_{TC}$ (tail complement) area. 
It seems that $C_{TC}$ has the same distribution as $C_T$ for all samples except for the sample "Productive". 
For latter sample $C_{TC}$ has slightly higher values than $C_T$. 
Note, also, that the "Productive" distribution has lower values for \hi than "Top h". 
This means that the height of the $C_{TC}$ areas is smaller for the "Productive" authors than for "Top H' ones. 
The previous remarks two lead us to the (rather expected) conclusion that the "Productive" authors have long and slim tails.

$C_{IC}$ distribution is shown in Figures~\ref{y0_c0_s1_distr_C_I_} and \ref{y0_c0_s1_distr_C_I__grouped}. 
In these plots, it is clear that the "Productive" authors have clearly higher values than any other sample since $C_IC$ is strongly related with the total number of publications.
%In these plots "Top h" and "Productive" seem to have exactly the same distribution. 
%Thus, this concludes that $C_{I}$ is highly correlated with the number of publications.
%Of course it is expected that "Random" and "Mortals" samples would have low values with lightly higher the "Mortals" one than the "Random".

%%%%- Areas C_P
%$C_P$ distributions (Figures~\ref{y0_c0_s1_distr_C_P_} and \ref{y0_c0_s1_distr_C_P__grouped}) are as expected similar to $C$ ones (Figure~\ref{y0_c0_s1_distr_p_} and \ref{y0_c0_s1_distr_p__grouped}).
%We also should compare these plots with the ones of $C_{I}$. 
%"Top h" and "Productive" have the same distribution for $C_{I}$ but not for $C_P$. 
%This means that we expect higher values for $PI$ for "Top h" authors than for "Productive".
%
%The distributions of $C_M$ (Figures~\ref{y0_c0_s1_distr_C_M_} and \ref{y0_c0_s1_distr_C_M__grouped}) show the big difference of "Top h" sample with the other ones with definitely higher values.
%%%%%%%%%%%%
\begin{figure}[!htb]
\centering	
%% 13
\subfigure[$PT$ (* 10000)]{\label{y0_c0_s1_distr_PT_} \resizebox{5.5cm}{3.8cm}{ \psfig{figure=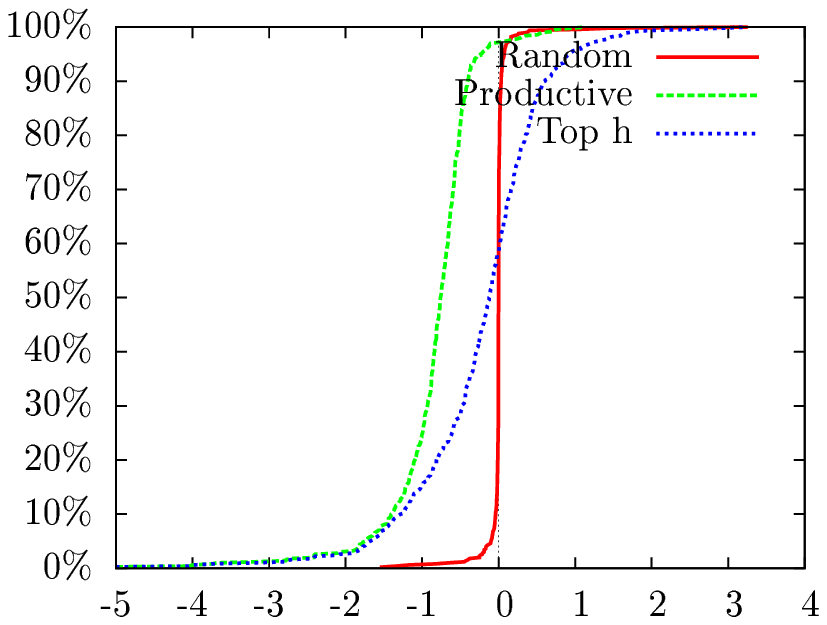,clip=}}}
\subfigure[$PT$ (* 10000)]{\label{y0_c0_s1_distr_PT__grouped} \resizebox{5.5cm}{3.8cm}{ \psfig{figure=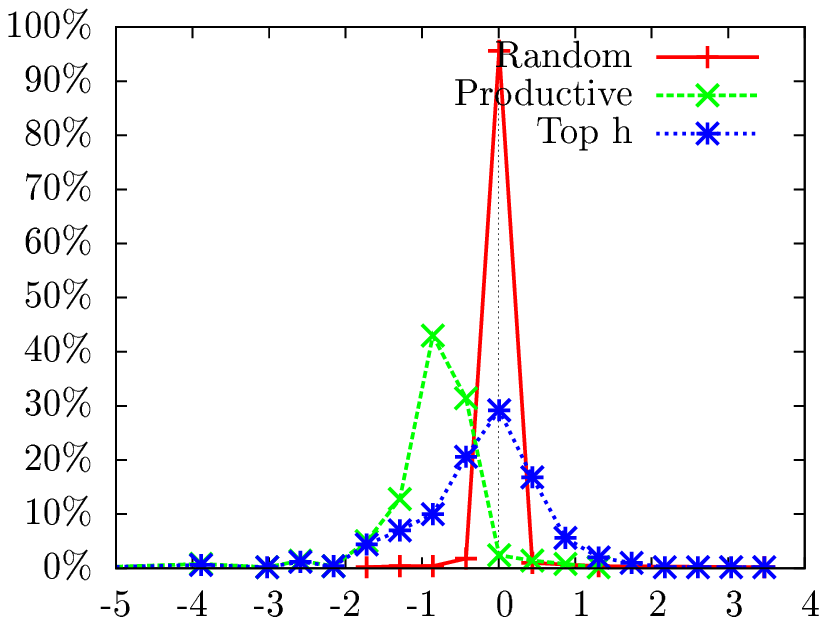,clip=}}}
%% 14
\subfigure[$PT_{(\kappa=2)}$ (* 10000)]{\label{y0_c0_s1_distr_PT_k2_} \resizebox{5.5cm}{3.8cm}{ \psfig{figure=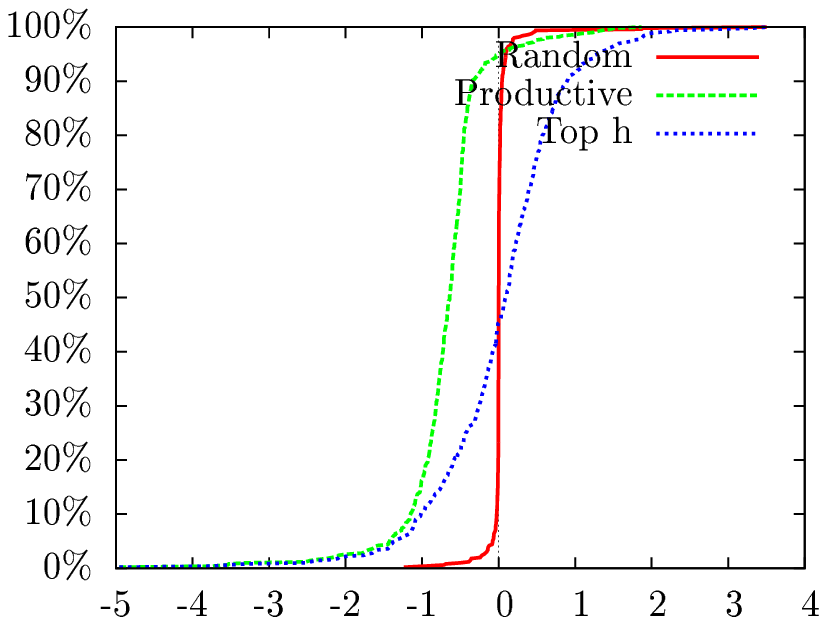,clip=}}}
\subfigure[$PT_{(\kappa=2)}$ (* 10000)]{\label{y0_c0_s1_distr_PT_k2__grouped} \resizebox{5.5cm}{3.8cm}{ \psfig{figure=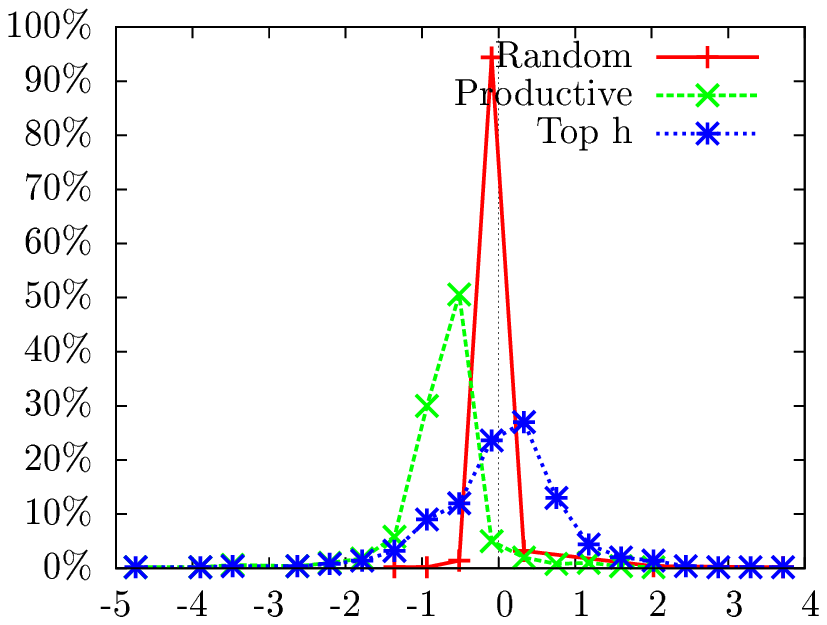,clip=}}}
	%% 18
\subfigure[$PT_{\kappa=4}$ (* 10000)]{\label{y0_c0_s1_distr_PT_k4_} \resizebox{5.5cm}{3.8cm}{ \psfig{figure=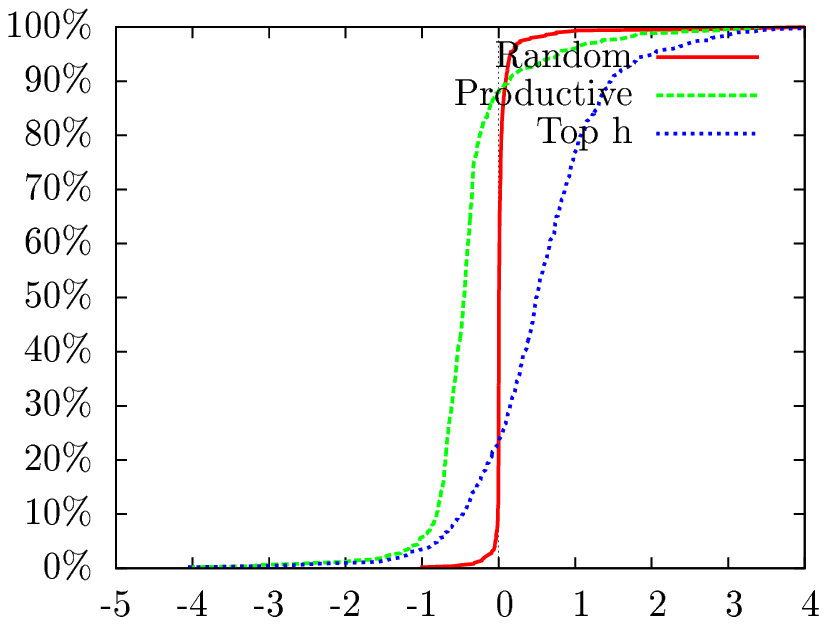,clip=}}}
\subfigure[$PT_{\kappa=4}$ (* 10000)]{\label{y0_c0_s1_distr_PT_k4__grouped} \resizebox{5.5cm}{3.8cm}{ \psfig{figure=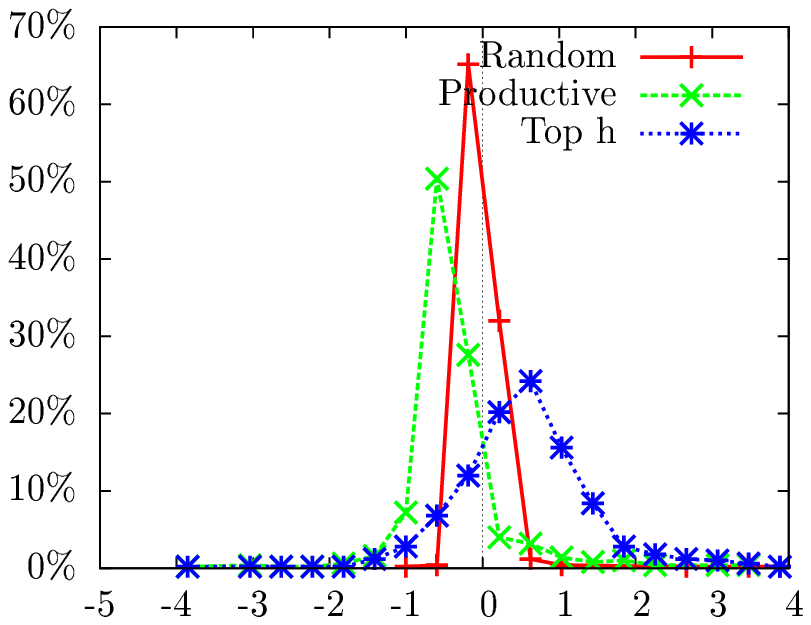,clip=}}}
%% 13 Range
\subfigure[$PT$ (limited x range)]{\label{y0_c0_s1_distr_PT__-500_500} \resizebox{5.5cm}{3.8cm}{ \psfig{figure=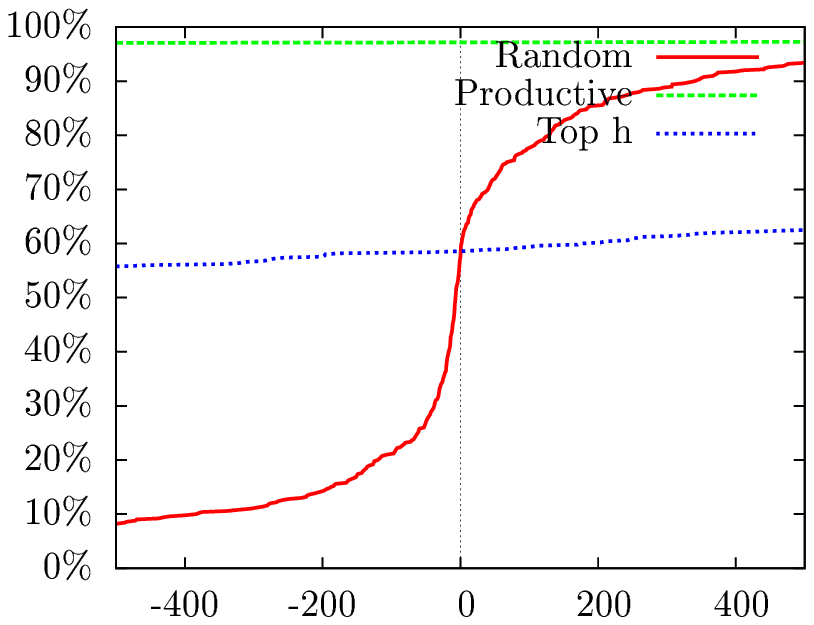,clip=}}}
%% 18 Range
\subfigure[$PT_{\kappa=4}$ (limited x range)]{\label{y0_c0_s1_distr_PT_k4__-500_500} \resizebox{5.5cm}{3.8cm}{ \psfig{figure=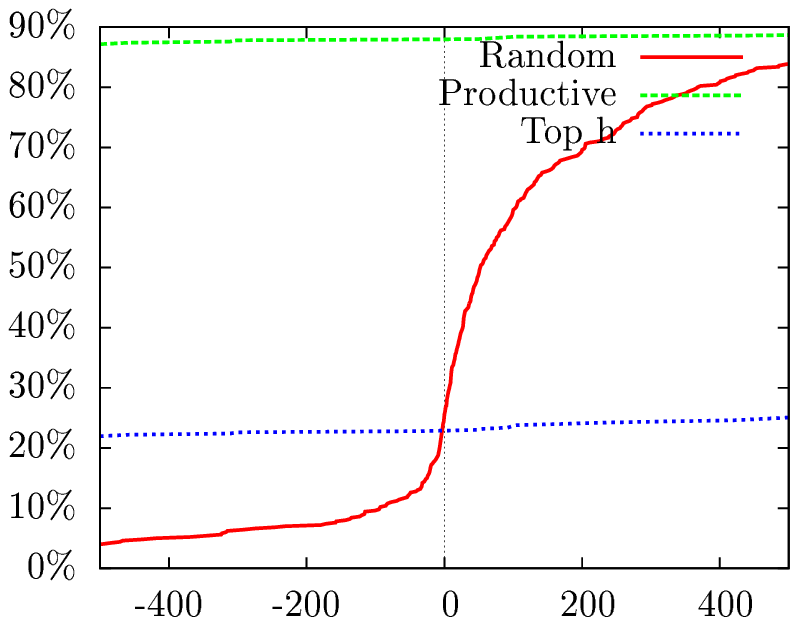,clip=}}}
\caption{Distributions of $PT$, $PT_{(\kappa=2)}$ and $PT_{(\kappa=4)}$}
\label{fig_plot_all_plot_sh_vals1}
\end{figure}
%\begin{figure}[ht]
%\centering
%	%% 15
%\subfigure[$PT_{(d=0)} = h^2-C_X$ (* 10000)]{\label{y0_c0_s1_distr_hindexC42hindex-C_X_} \resizebox{5.5cm}{3.8cm}{ \psfig{figure=plots/density_all/y0_c0_s1_distr_hindexC42hindex-C_X_.eps,clip=}}}
%\subfigure[$PT_{(d=0)} = h^2-C_X$ (* 10000)]{\label{y0_c0_s1_distr_hindexC42hindex-C_X__grouped} \resizebox{5.5cm}{3.8cm}{ \psfig{figure=plots/density_all/y0_c0_s1_distr_hindexC42hindex-C_X__grouped.eps,clip=}}}
%%% 15 Range
%\subfigure[$PT_{(d=0)} = h^2-C_X$ (limited x range)]{\label{y0_c0_s1_distr_hindexC42hindex-C_X__-500_500} \resizebox{5.5cm}{3.8cm}{ \psfig{figure=plots/density_all/y0_c0_s1_distr_hindexC42hindex-C_X__-500_500.eps,clip=}}}
%\caption{Distributions of $PT_{(d=0)}$}
%\label{fig_plot_all_plot_sh_vals2}
%\end{figure}

\begin{figure}[ht]
\centering
%% 18
\subfigure[$PI$ (* 100000)]{\label{y0_c0_s1_distr_PI_} \resizebox{5.5cm}{3.8cm}{ \psfig{figure=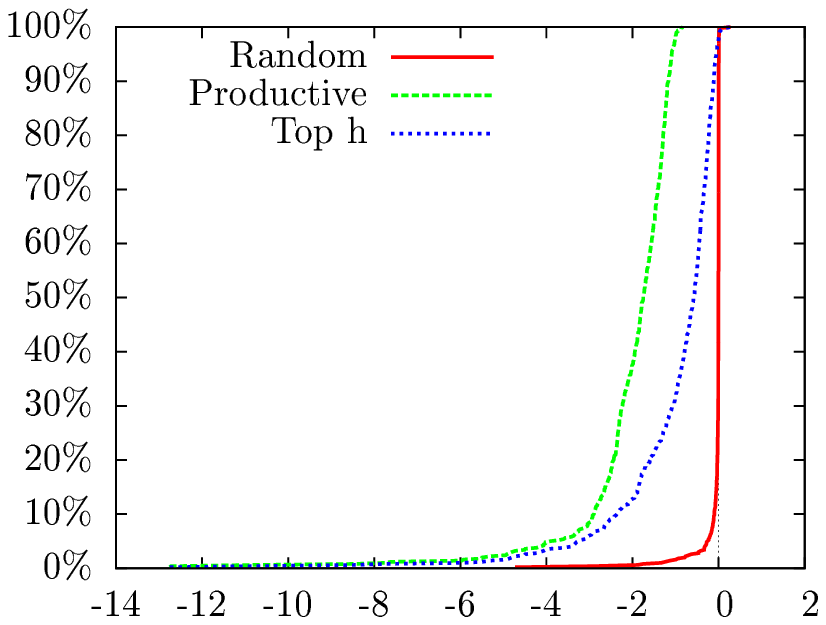,clip=}}}
\subfigure[$PI$ (* 100000)]{\label{y0_c0_s1_distr_PI__grouped} \resizebox{5.5cm}{3.8cm}{ \psfig{figure=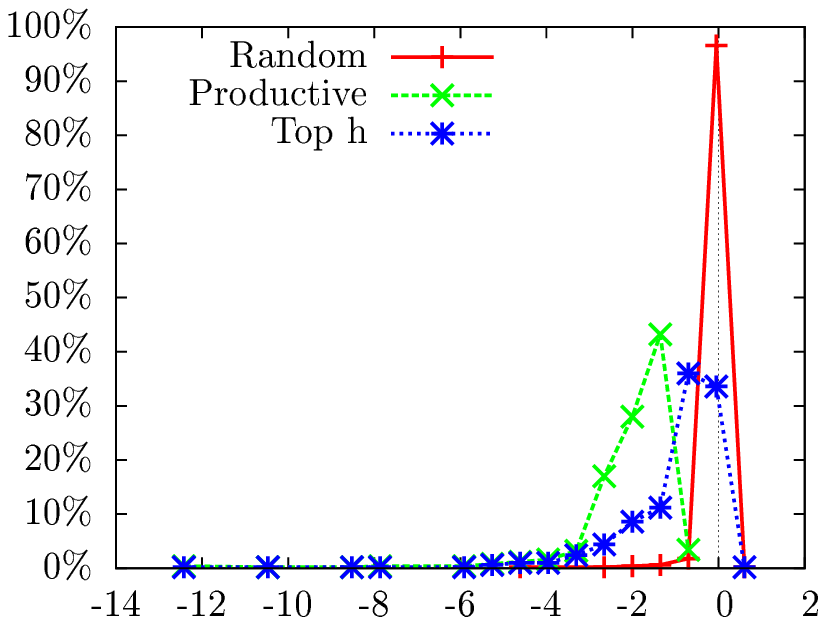,clip=}}}

%% 18 Range
\subfigure[$PI$ (limited x range)]{\label{y0_c0_s1_distr_PI__-500_500} \resizebox{5.5cm}{3.8cm}{ \psfig{figure=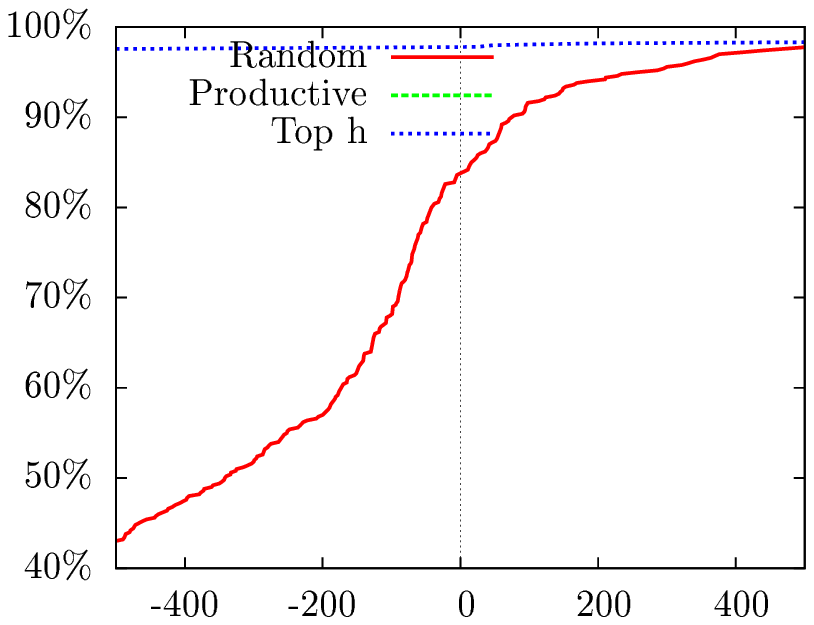,clip=}}}
\caption{Distributions of $PI$}
\label{fig_plot_all_plot_vsh_vals1}
\end{figure}

%\begin{figure}[ht]
%\centering	
%%% 18
%\subfigure[$PI_{(d=0)} = h^2-C_{I}$ (* 100000)]{\label{y0_c0_s1_distr_hindexC42hindex-C_I_} \resizebox{5.5cm}{3.8cm}{ \psfig{figure=plots/density_all/y0_c0_s1_distr_hindexC42hindex-C_I_.eps,clip=}}}
%\subfigure[$PI_{(d=0)} = h^2-C_{I}$ (* 100000)]{\label{y0_c0_s1_distr_hindexC42hindex-C_I__grouped} \resizebox{5.5cm}{3.8cm}{ \psfig{figure=plots/density_all/y0_c0_s1_distr_hindexC42hindex-C_I__grouped.eps,clip=}}}
%%% 18 Range
%\subfigure[$PI_{(d=0)} = h^2-C_{I}$ (limited x range)]{\label{y0_c0_s1_distr_hindexC42hindex-C_I__-500_500} \resizebox{5.5cm}{3.8cm}{ \psfig{figure=plots/density_all/y0_c0_s1_distr_hindexC42hindex-C_I__-500_500.eps,clip=}}}
%\caption{Distributions of $PI_{(d=0)}$ }
%\label{fig_plot_all_plot_vsh_vals2}
%\end{figure}

In Figure \ref{fig_plot_all_plot_sh_vals1} we see the distributions for the previously defined $PT$ index. 
Interestingly, it seems that the value 0 is a key value. 
For all plots, the zero y-axis is the center of the figure.
As seen in Figures \ref{y0_c0_s1_distr_PT_} and \ref{y0_c0_s1_distr_PT__grouped} most of the authors are located around zero. 
Note that in the right plots, a point at $x=0, y=95\%$ with a previous xtic at $x=-3000$ means that the 95\% of the authors have values between in the range -1500..1500. 
The first two plots show that the "Top h" authors have the highest values for $PT$ (about 10\% of them have values greater than 8000). 

Figure ~\ref{y0_c0_s1_distr_PT__-500_500} is a zoomed version of Figure \ref{y0_c0_s1_distr_PT_}. 
In this figure, it is clear that about 96\% of the "Productive" authors have $PT<0$.
This means that in this sample there are a lot of "mass producers" (people with high number of publications but relatively low \hi - or at least not in "Top h-indexers").
The other samples cut the zero y-axis at about 50\% to 60\%, which means that 40\% to 50\% are positive.
% The "mortals" line is placed between the "Top h" and "Random" one, which is rather expected. 
It is also noticeable that about 70\% (15-85\%) of the "Random sample have values very close to zero within the range -200..200.

%%%%%%------------here
In the Figure (\ref{y0_c0_s1_distr_PT_k2_}, \subref{y0_c0_s1_distr_PT_k2__grouped}, \subref{y0_c0_s1_distr_PT_k4_} and \subref{y0_c0_s1_distr_PT_k4__grouped}) we also present the distributions for $PT_{\kappa=2}$ and $PT_{\kappa=4}$.
We remind that factor $\kappa$ is the core area multiplier.
In these plots, it is shown that these distributions behave like the basic $PT$ distribution except that they are slightly shifted to the right.
The "Top h" sample is affected more than the others.
This outcome is understood since they are the authors with the greatest \hi values; in other words, they have the greatest \hi core areas.

Comparing subfigure \subref{y0_c0_s1_distr_PT_k4__-500_500} to \subref{y0_c0_s1_distr_PT__-500_500} we can better visualize the differences.
The number of authors in the negative side of samples "Random" and "Productive" have been decreased from 70\% to 25\%, meaning that 45\% of the sample members moved from the negative to the positive side.
The number of "Productive" authors in the negative side has been decreased from 95\% to 80\%, i.e. an additional 15\% of the sample members moved to the positive side.
% Finally, 30\% (70\% minus 40\%) of the "Mortal" sample are moving to the positive side by setting $\kappa=4$ instead of the default $\kappa=1$.

In addition to the distribution plots, Table~\ref{positive_tab} presents the number of authors that have the mentioned metrics below or above zero for each sample.
As mentioned before, 97\% of the "Productive" authors have $PT<0$, whereas only 3\% reside in the positive side of the plot.
This amount increases as we increase the core factor $\kappa$. For $\kappa=4$ the increment is 17\% (21\% from 4\%).
In all other samples the increment is greater, i.e. for "Top h" the increment is 33\%, for "Random" is 31\%.

In Figure \ref{fig_plot_all_plot_vsh_vals1} the same kind of plots are presented for the metric $PI$. 
The difference is, as expected, that most of the authors lie in the negative side of the graph.
The cut points of y-axis are also presented in Table~\ref{positive_tab}.
About 2\% of the "Top \hi" authors have $PI>0$ but {\bf none} of the "Productive" authors.
The cut point for "Random" authors is at 6\%. 
%whereas for "Mortals" it is 7\%. 
Also, at this point we repeat the experiment of varying the $\kappa$ value.
The results do not match with those of $PT$ case.
Incrementing $\kappa$ does not increase the number of positive authors in the same way as the $PT$ case.
The increment is negligible for the "Productive" and "Top h" and very small for the sample "Random".
% and "Mortals".
This leads to the conclusion that varying the $\kappa$ factor does not affect $PI$ significantly. 
Probably different default values for the factors of Equation \ref{eq_PI} (especially for $\kappa$ and/or $\iota$ ) may be needed for tuning the $PI$ metric. 
However, this task remains out of the scope of the present article.

\begin{table}[!bt]
\centering
\footnotesize 
\begin{tabular}{|r@{\hspace{0.3em}}|@{\hspace{0.3em}}cc@{\hspace{0.3em}}|@{\hspace{0.3em}}cc@{\hspace{0.3em}}|@{\hspace{0.3em}}cc@{\hspace{0.3em}}|@{\hspace{0.3em}}cc@{\hspace{0.3em}}|@{\hspace{0.3em}}cc@{\hspace{0.3em}}|@{\hspace{0.3em}}cc@{\hspace{0.3em}}|}\hline
\multirow{2}{*}{\vspace{0.1cm} Sample|} &\multicolumn{2}{@{\hspace{-0.3em}}|c@{\hspace{-0.3em}}}{$PT$} &\multicolumn{2}{@{\hspace{-0.3em}}|c}{$PT_{\kappa=2}$} &\multicolumn{2}{@{\hspace{-0.3em}}|c}{$PT_{\kappa=4}$} &\multicolumn{2}{@{\hspace{-0.3em}}|c}{$PI$} &\multicolumn{2}{@{\hspace{-0.3em}}|c}{$PI_{\kappa=2}$} &\multicolumn{2}{@{\hspace{-0.3em}}|c|}{$PI_{\kappa=4}$}\\
 & $<0$ & $\ge 0$ & $<0$ & $\ge 0$ & $<0$ & $\ge 0$ & $<0$ & $\ge 0$ & $<0$ & $\ge 0$ & $<0$ & $\ge 0$\\ \hline\hline
%\multirow{2}{*}{Random|} & 631 & 322 & 529 & 424 & 329 & 624 & 885 & 68 & 860 & 93 & 833 & 120\\
% & 66\% &34\% & 56\% &44\% & 35\% &65\% & 93\% &7\% & 90\% &10\% & 87\% &13\%\\\hline
%\multirow{2}{*}{Productive|} & 418 & 18 & 399 & 37 & 343 & 93 & 435 & 1 & 435 & 1 & 435 & 1\\
% & 96\% &4\% & 92\% &8\% & 79\% &21\% & 100\% &0\% & 100\% &0\% & 100\% &0\%\\\hline
%\multirow{2}{*}{Top $h$|} & 242 & 158 & 192 & 208 & 107 & 293 & 392 & 8 & 391 & 9 & 385 & 15\\
% & 60\% &40\% & 48\% &52\% & 27\% &73\% & 98\% &2\% & 98\% &2\% & 96\% &4\%\\\hline
%\multirow{2}{*}{Mortals|} & 280 & 120 & 232 & 168 & 156 & 244 & 374 & 26 & 365 & 35 & 359 & 41\\
% & 70\% &30\% & 58\% &42\% & 39\% &61\% & 94\% &6\% & 91\% &9\% & 90\% &10\%\\\hline
%\multirow{2}{*}{Unioned|} & 1509 & 603 & 1301 & 811 & 904 & 1208 & 2010 & 102 & 1975 & 137 & 1936 & 176\\
% & 71\% &29\% & 62\% &38\% & 43\% &57\% & 95\% &5\% & 94\% &6\% & 92\% &8\%\\\hline

\multirow{2}{*}{Random} & 284 & 216 & 213 & 287 & 122 & 378 & 418 & 82 & 408 & 92 & 383 & 117\\
 & 57\% &43\% & 43\% &57\% & 24\% &76\% & 84\% &16\% & 82\% &18\% & 77\% &23\%\\\hline
\multirow{2}{*}{Productive} & 485 & 15 & 474 & 26 & 439 & 61 & 500 & 0 & 500 & 0 & 500 & 0\\
 & 97\% &3\% & 95\% &5\% & 88\% &12\% & 100\% &0\% & 100\% &0\% & 100\% &0\%\\\hline
\multirow{2}{*}{Top $h$} & 292 & 208 & 226 & 274 & 114 & 386 & 488 & 12 & 484 & 16 & 473 & 27\\
 & 58\% &42\% & 45\% &55\% & 23\% &77\% & 98\% &2\% & 97\% &3\% & 95\% &5\%\\\hline
\multirow{2}{*}{Unioned} & 904 & 419 & 767 & 556 & 563 & 760 & 1230 & 93 & 1216 & 107 & 1180 & 143\\
 & 68\% &32\% & 58\% &42\% & 43\% &57\% & 93\% &7\% & 92\% &8\% & 89\% &11\%\\\hline
\end{tabular}
\normalsize
\caption{$PT$ and $PI$ statistics}
\label{positive_tab}
\end{table}

%\subsection{Experimental Comparisons}
%\label{sec:comps}

\subsection{$PT$ robustness to self-citations}
\label{subsec:self}

\begin{figure}[!tb]
\centering
\subfigure[$h$ vs. $h$(no-self)]{\label{qqplots_hindex_c0_s1_y0_vs_hindex_c0_s0_y0_all} \resizebox{5.5cm}{3.8cm}{ \psfig{figure=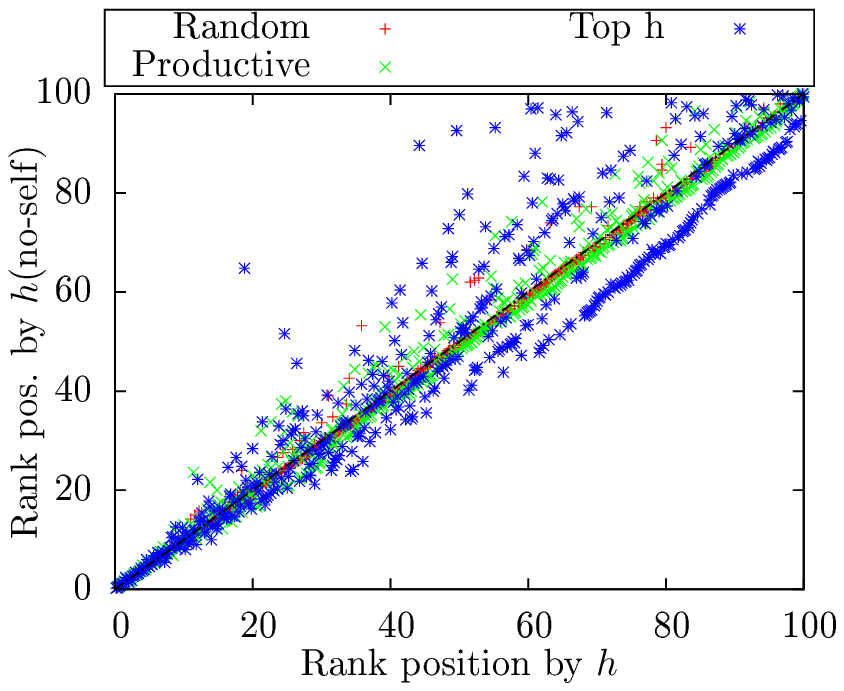,clip=}}}
%% 
% \subfigure[$C_S$ vs. $C_S$(no-self)]{\label{qqplots_C_S_c0_s1_y0_vs_C_S_c0_s0_y0_all} \resizebox{5.5cm}{3.8cm}{ % % \psfig{figure=plots/qqplots/qqplots_C_S_c0_s1_y0_vs_C_S_c0_s0_y0_all.eps,clip=}}}
%% 
\subfigure[$PT$ vs. $PT$(no-self)]{\label{qqplots_PT_c0_s1_y0_vs_PT_c0_s0_y0_all} \resizebox{5.5cm}{3.8cm}{ \psfig{figure=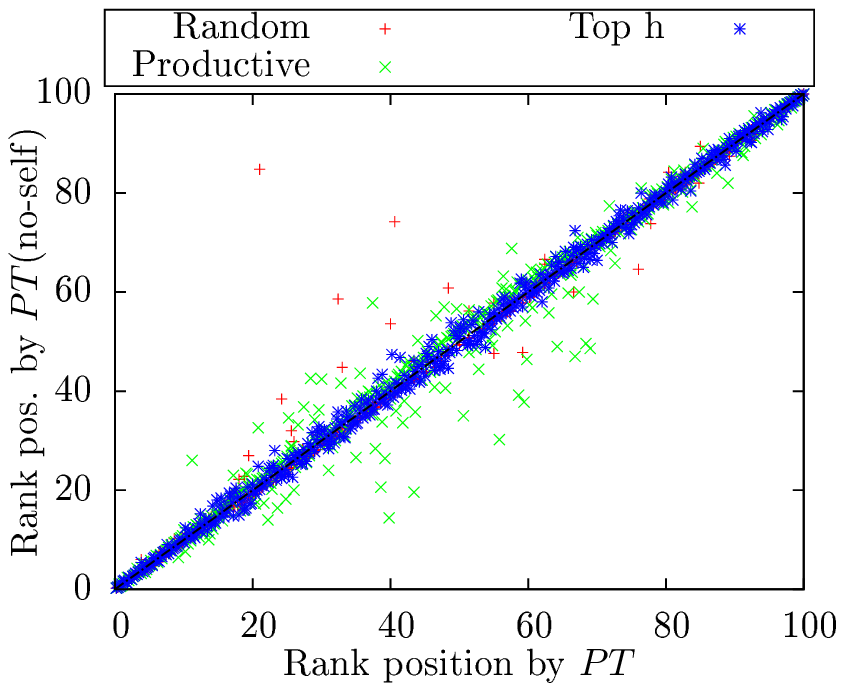,clip=}}}
%% 
%\subfigure[$PI$ vs. $PI$(no-self)]{\label{qqplots_PI_c0_s1_y0_vs_PI_c0_s0_y0_all} \resizebox{5.5cm}{3.8cm}{ \psfig{figure=plots/qqplots/qqplots_PI_c0_s1_y0_vs_PI_c0_s0_y0_all.eps,clip=}}}
\caption{Q-Q plots: X and Y axis denote the rank position normalized in percent.}
\label{plot_all_qqplot_self}
\end{figure}

% If we do not take care about self-citations, then the rank results may be unfair. 
We performed another parallel experiment to study the behavior of the new metrics with respect to self-citations. 
A citation is considered as self if there is at least one common author between the citing and the cited paper.
In Figure~\ref{qqplots_hindex_c0_s1_y0_vs_hindex_c0_s0_y0_all} a qq-plot is shown, which compares the ranking produced by \hi. 
The x-axis represents the rank produced by the computed \hi including self-citations, whereas y-axis represents the rank of \hi after excluding self-citations.
%It is remarked that the rank order does not differ significantly.
We have performed several experiments with different types of ranking and they all show similar behavior with respect to \hi.

%In Figure~\ref{qqplots_C_S_c0_s1_y0_vs_C_S_c0_s0_y0_all} the rank comparison between "rank by total number of citations" and "rank by total number of citations excluding self-ones" is presented.
%It seems that the \hi ranking and the "times cited" ranking are both affected from the self-citations. 
%Also they exhibit similar behavior.

In Figure \ref{qqplots_PT_c0_s1_y0_vs_PT_c0_s0_y0_all} the same kind of qq-plot for the $PT$ as a rank criterion is displayed.
It is apparent that $PT$ is much less affected by self-citations than the \hi. 
%Also, $PI$ is almost not affected.
%Thus, although $PI$ produces in the vast majority negative numbers, it could be used for self-citation unaffected ranking.
This is another advantage of our new metrics; they are not affected by self-citations as other metrics and, thus, someone can rely more safely on them.

% \input{inc_tex/plots_noself.tex}
% In Figure \ref{plot_all_plot_self0} are presented the distribution plots of \hi, $C$, $PT$ and $PI$, all computed by excluding the self-citations. 
% It seems that there are no significant differences from the plots that include the self-citations except that the values are lightly smaller.

\section{Rank Results}
\label{sec:results}

In this section we present some rank tables produced by using our data extracted from MAS.
Table \ref{tab_rank_table_hindex} shows the rank table for top-20 authors by \hi from all our samples.
The corresponding $PT$ values are also shown, as well as their rank positions according to $PT$. 
It is remarkable that about half of them are characterized as 
"Mass Producers" as the have negative $PT$ values. 
%This is due to the fact that most of them have a lot of publications.

\begin{table}[!bt]
\begin{center}
\footnotesize

\begin{tabular}{|l@{}||@{\hspace{.1cm}}rr@{\hspace{.1cm}}|@{\hspace{.1cm}}rr@{\hspace{.1cm}}|r|r|r|}\hline
\multirow{2}{*}{\bf Author } 	&\multicolumn{2}{@{\hspace{-.1cm}}|c@{\hspace{.1cm}}}{\bf $h$} 	&\multicolumn{2}{@{\hspace{-.1cm}}|c@{\hspace{.1cm}}|}{\bf $PT$} & \multicolumn{1}{|c|}{\multirow{2}{*}{\bf $p$}}& \multicolumn{1}{|c|}{\multirow{2}{*}{\bf $C$}}& \multicolumn{1}{|c|}{\multirow{2}{*}{\bf $C/p$}}\\
	&val	&pos	&val	&pos &  &  & \\\hline\hline
Shenker Scott		& 97 & 1	& 5754 	& 52   	& 508 		& 45621 		& 89.81 	\\ 
Foster Ian		& 93 & 2	& -15510 	& 1287   	& 768 		& 47265 		& 61.54 	\\ 
Garcia-molina Hector		& 92 & 3	& -17423 	& 1299   	& 605 		& 29773 		& 49.21 	\\ 
Estrin Deborah		& 90 & 4	& 5348 	& 62   	& 479 		& 40358 		& 84.25 	\\ 
Ullman Jeffrey		& 86 & 5	& 11267 	& 18   	& 460 		& 43431 		& 94.42 	\\ 
Culler David		& 84 & 6	& 7552 	& 38   	& 386 		& 32920 		& 85.28 	\\ 
Tarjan Robert		& 83 & 7	& 2888 	& 117   	& 405 		& 29614 		& 73.12 	\\ 
Towsley Don		& 82 & 8	& -31929 	& 1318   	& 793 		& 26373 		& 33.26 	\\ 
Kanade T.		& 81 & 9	& -20753 	& 1309   	& 742 		& 32788 		& 44.19 	\\ 
Haussler David		& 81 & 10	& 10952 	& 19   	& 335 		& 31526 		& 94.11 	\\ 
Jain Anil		& 81 & 11	& -11474 	& 1236   	& 590 		& 29755 		& 50.43 	\\ 
Papadimitriou Christos		& 80 & 12	& -5897 	& 968   	& 506 		& 28183 		& 55.70 	\\ 
Katz Randy		& 78 & 13	& -27820 	& 1317   	& 757 		& 25142 		& 33.21 	\\ 
Pentland Alex		& 77 & 14	& -1242 	& 724   	& 509 		& 32022 		& 62.91 	\\ 
Han Jiawei		& 77 & 15	& -15410 	& 1285   	& 653 		& 28942 		& 44.32 	\\ 
Jordan Michael		& 75 & 16	& -1062 	& 717   	& 499 		& 30738 		& 61.60 	\\ 
Karp Richard		& 75 & 17	& 7231 	& 41   	& 377 		& 29881 		& 79.26 	\\ 
Zisserman A.		& 75 & 18	& 210 	& 263   	& 421 		& 26160 		& 62.14 	\\ 
Jennings Nick		& 74 & 19	& -15718 	& 1289   	& 626 		& 25130 		& 40.14 	\\ 
Thrun S.		& 74 & 20	& -5789 	& 958   	& 445 		& 21665 		& 48.69 	\\ 

\hline
\end{tabular}
\normalsize
\caption{Rank table by $h$ (top-20 by \hi)}
\label{tab_rank_table_hindex}
\end{center}
\end{table}

\begin{table}[!bt]
\begin{center}
\footnotesize

\begin{tabular}{|l@{}||@{\hspace{.1cm}}rr@{\hspace{.1cm}}|@{\hspace{.1cm}}rr@{\hspace{.1cm}}|r|r|r|}\hline
\multirow{2}{*}{\bf Author } 	&\multicolumn{2}{@{\hspace{-.1cm}}|c@{\hspace{.1cm}}}{\bf $PT$} 	&\multicolumn{2}{@{\hspace{-.1cm}}|c@{\hspace{.1cm}}|}{\bf $h$} & \multicolumn{1}{|c|}{\multirow{2}{*}{\bf $p$}}& \multicolumn{1}{|c|}{\multirow{2}{*}{\bf $C$}}& \multicolumn{1}{|c|}{\multirow{2}{*}{\bf $C/p$}}\\
	&val	&pos	&val	&pos &  &  & \\\hline\hline
Vapnik Vladimir		& 32542 & 1	& 50 	& 171   	& 126 		& 36342 		& 288.43 	\\ 
Rivest Ronald		& 29340 & 2	& 62 	& 53   	& 320 		& 45336 		& 141.68 	\\ 
Zadeh L.		& 25613 & 3	& 59 	& 70   	& 320 		& 41012 		& 128.16 	\\ 
Kohonen Teuvo		& 19880 & 4	& 51 	& 157   	& 160 		& 25439 		& 158.99 	\\ 
Floyd Sally		& 18059 & 5	& 66 	& 38   	& 222 		& 28355 		& 127.73 	\\ 
Kesselman Carl		& 17054 & 6	& 60 	& 64   	& 272 		& 29774 		& 109.46 	\\ 
Schapire Robert		& 16169 & 7	& 56 	& 90   	& 186 		& 23449 		& 126.07 	\\ 
Milner Robin		& 16019 & 8	& 54 	& 108   	& 202 		& 24011 		& 118.87 	\\ 
Shamir A		& 15926 & 9	& 53 	& 125   	& 213 		& 24406 		& 114.58 	\\ 
Tuecke Steven		& 14747 & 10	& 44 	& 281   	& 96 		& 17035 		& 177.45 	\\ 
Balakrishnan Hari		& 14444 & 11	& 72 	& 21   	& 272 		& 28844 		& 106.04 	\\ 
Agrawal Rakesh		& 14375 & 12	& 67 	& 30   	& 353 		& 33537 		& 95.01 	\\ 
Hinton G.		& 13415 & 13	& 63 	& 45   	& 314 		& 29228 		& 93.08 	\\ 
Aho Alfred		& 13048 & 14	& 50 	& 173   	& 193 		& 20198 		& 104.65 	\\ 
Lamport Leslie		& 12254 & 15	& 59 	& 71   	& 273 		& 24880 		& 91.14 	\\ 
Hopcroft John		& 12088 & 16	& 45 	& 258   	& 198 		& 18973 		& 95.82 	\\ 
Morris Robert		& 11685 & 17	& 57 	& 81   	& 305 		& 25821 		& 84.66 	\\ 
Ullman Jeffrey		& 11267 & 18	& 86 	& 5   	& 460 		& 43431 		& 94.42 	\\ 
Haussler David		& 10952 & 19	& 81 	& 10   	& 335 		& 31526 		& 94.11 	\\ 
Joachims T.		& 10767 & 20	& 41 	& 377   	& 134 		& 14580 		& 108.81 	\\ 

\hline
\end{tabular}
\normalsize
\caption{Rank table by $PT$ (top-20 influential)}
\label{tab_rank_table_PT}
\end{center}
\end{table}

\begin{table}[!bt]
\begin{center}
\footnotesize

\begin{tabular}{|l@{}||@{\hspace{.1cm}}rr@{\hspace{.1cm}}|@{\hspace{.1cm}}rr@{\hspace{.1cm}}|r|r|r|}\hline
\multirow{2}{*}{\bf Author } 	&\multicolumn{2}{@{\hspace{-.1cm}}|c@{\hspace{.1cm}}}{\bf $PT$} 	&\multicolumn{2}{@{\hspace{-.1cm}}|c@{\hspace{.1cm}}|}{\bf $h$} & \multicolumn{1}{|c|}{\multirow{2}{*}{\bf $p$}}& \multicolumn{1}{|c|}{\multirow{2}{*}{\bf $C$}}& \multicolumn{1}{|c|}{\multirow{2}{*}{\bf $C/p$}}\\
	&val	&pos	&val	&pos &  &  & \\\hline\hline
Ikeuchi Katsushi		& -18173 & 1303	& 43 	& 327   	& 638 		& 7412 		& 11.62 	\\ 
Thalmann D.		& -18356 & 1304	& 46 	& 249   	& 632 		& 8600 		& 13.61 	\\ 
Reddy Sudhakar		& -18369 & 1305	& 43 	& 322   	& 659 		& 8119 		& 12.32 	\\ 
Gao Wen		& -18494 & 1306	& 26 	& 649   	& 907 		& 4412 		& 4.86 	\\ 
Prade Henri		& -18692 & 1307	& 65 	& 42   	& 633 		& 18228 		& 28.80 	\\ 
Liu K.		& -19063 & 1308	& 42 	& 355   	& 672 		& 7397 		& 11.01 	\\ 
Kanade T.		& -20753 & 1309	& 81 	& 9   	& 742 		& 32788 		& 44.19 	\\ 
Rosenfeld Azriel		& -21023 & 1310	& 59 	& 73   	& 707 		& 17209 		& 24.34 	\\ 
Gupta Anoop		& -23959 & 1311	& 64 	& 43   	& 739 		& 19241 		& 26.04 	\\ 
Miller J.		& -24112 & 1312	& 40 	& 433   	& 807 		& 6568 		& 8.14 	\\ 
Shin Kang		& -24125 & 1313	& 57 	& 85   	& 731 		& 14293 		& 19.55 	\\ 
Schmidt Douglas		& -24153 & 1314	& 56 	& 94   	& 729 		& 13535 		& 18.57 	\\ 
Bertino Elisa		& -27058 & 1315	& 49 	& 194   	& 805 		& 9986 		& 12.40 	\\ 
Yu Philip		& -27727 & 1316	& 63 	& 48   	& 789 		& 18011 		& 22.83 	\\ 
Katz Randy		& -27820 & 1317	& 78 	& 13   	& 757 		& 25142 		& 33.21 	\\ 
Towsley Don		& -31929 & 1318	& 82 	& 8   	& 793 		& 26373 		& 33.26 	\\ 
Kuo C.		& -36848 & 1319	& 40 	& 425   	& 1148 		& 7472 		& 6.51 	\\ 
GERLA MARIO		& -37464 & 1320	& 67 	& 32   	& 945 		& 21362 		& 22.61 	\\ 
Dongarra Jack		& -39901 & 1321	& 67 	& 31   	& 982 		& 21404 		& 21.80 	\\ 
Poor H.		& -40492 & 1322	& 55 	& 100   	& 1069 		& 15278 		& 14.29 	\\ 
Huang Thomas		& -54047 & 1323	& 67 	& 33   	& 1172 		& 19988 		& 17.05 	\\ 

\hline
\end{tabular}
\normalsize
\caption{Rank table by $PT$ (Last-20 - Top-20 Mass Producers)}
\label{tab_rank_table_PT_last20}
\end{center}
\end{table}

It Table \ref{tab_rank_table_PT} we show the rank list ordered by $PT$; all authors have high rank position by \hi as well. 
For all of them it holds that $h>44$ except for one author, Kennedy James, who has \hi equal to 20. 
It can be seen that Kennedy James has only 35 publications but a total number of citations of 15482.
Also, all authors that appear in this top-20 list have less than 350 publications, except for Ullman J. who has 460 with a total number of citations of 43431.
All authors have a huge number of citations. 
For example, Rivest Ronald is ranked second by $PT$ with 45344 citations, as much as Shenker S. who is ranked first by \hi. 
This fact despite that Rivest Ronald has 321 publications, whereas Shenker S. has 508. 
Apparently, Rivest Ronald has a better citation curve. 
That is why Rivest Ronald chimps from the 63th place by \hi to the 2nd by $PT$ and overtakes Shenker who was first by \hi.
% Shenker S. falls from the 1st place by \hi to the 55th by $PT$.

Based on the above rank tables, someone could say that $PT$ is correlated to the average number of citations per publication.
Figures \ref{qqplots_citp_c0_s1_y0_vs_PT_c0_s1_y0_union} and \ref{qqplots_citp_c0_s1_y0_vs_hindex_c0_s1_y0_union} show that this is not the case; actually \hi is much more correlated to $C/p$ than $PT$ (closer to line $x=y$).

Table \ref{tab_rank_table_PT_last20} shows the top-20 "Mass Producers" from our samples. 
In this table we also present the average number of citations per paper ($C/p$ column).
It can be seen that there is a big range of average values from 4 to 45 citations per publication in the top "Mass Producers".

\begin{table}[!bt]
\begin{center}
\footnotesize

\begin{tabular}{|l@{}||@{\hspace{.1cm}}rr@{\hspace{.1cm}}|@{\hspace{.1cm}}rr@{\hspace{.1cm}}|r|r|r|}\hline
\multirow{2}{*}{\bf Author } 	&\multicolumn{2}{@{\hspace{-.1cm}}|c@{\hspace{.1cm}}}{\bf $PT$} 	&\multicolumn{2}{@{\hspace{-.1cm}}|c@{\hspace{.1cm}}|}{\bf $h$} & \multicolumn{1}{|c|}{\multirow{2}{*}{\bf $p$}}& \multicolumn{1}{|c|}{\multirow{2}{*}{\bf $C$}}& \multicolumn{1}{|c|}{\multirow{2}{*}{\bf $C/p$}}\\
	&val	&pos	&val	&pos &  &  & \\\hline\hline
Agrawal Rakesh		& 14375 & 1	& 67 	& 8   	& 353 		& 33537 		& 95.01 	\\ 
Ullman Jeffrey		& 11267 & 2	& 86 	& 2   	& 460 		& 43431 		& 94.42 	\\ 
Motwani Rajeev		& 9349 & 3	& 69 	& 6   	& 271 		& 23287 		& 85.93 	\\ 
Fagin Ronald		& 4400 & 4	& 59 	& 16   	& 215 		& 13604 		& 63.27 	\\ 
Widom Jennifer		& 4031 & 5	& 71 	& 4   	& 280 		& 18870 		& 67.39 	\\ 
Florescu Daniela		& 3058 & 6	& 40 	& 43   	& 132 		& 6738 		& 51.05 	\\ 
Bernstein Philip		& 2917 & 7	& 52 	& 22   	& 279 		& 14721 		& 52.76 	\\ 
Buneman Peter		& 2001 & 8	& 43 	& 39   	& 158 		& 6946 		& 43.96 	\\ 
Hellerstein Joseph		& 1941 & 9	& 51 	& 25   	& 272 		& 13212 		& 48.57 	\\ 
Naughton J.		& 640 & 10	& 48 	& 29   	& 221 		& 8944 		& 40.47 	\\ 
Dewitt David		& 308 & 11	& 63 	& 10   	& 308 		& 15743 		& 51.11 	\\ 
Koudas Nick		& 58 & 12	& 35 	& 50   	& 168 		& 4713 		& 28.05 	\\ 
Sagiv Yehoshua		& -196 & 13	& 42 	& 40   	& 209 		& 6818 		& 32.62 	\\ 
Chaudhuri Surajit		& -278 & 14	& 41 	& 41   	& 239 		& 7840 		& 32.80 	\\ 
Egenhofer Max		& -314 & 15	& 47 	& 30   	& 223 		& 7958 		& 35.69 	\\ 
Livny Miron		& -597 & 16	& 61 	& 12   	& 310 		& 14592 		& 47.07 	\\ 
Suciu Dan		& -659 & 17	& 54 	& 19   	& 285 		& 11815 		& 41.46 	\\ 
Papadias Dimitris		& -809 & 18	& 38 	& 47   	& 200 		& 5347 		& 26.73 	\\ 
Lakshmanan Laks		& -914 & 19	& 37 	& 48   	& 196 		& 4969 		& 25.35 	\\ 
Lenzerini M.		& -1074 & 20	& 50 	& 26   	& 269 		& 9876 		& 36.71 	\\ 
Abiteboul Serge		& -1111 & 21	& 59 	& 15   	& 321 		& 14347 		& 44.69 	\\ 
Ioannidis Yannis		& -1647 & 22	& 39 	& 46   	& 209 		& 4983 		& 23.84 	\\ 
Sellis Timos		& -2747 & 23	& 36 	& 49   	& 264 		& 5461 		& 20.69 	\\ 
Jagadish H.		& -2924 & 24	& 52 	& 24   	& 303 		& 10128 		& 33.43 	\\ 
Dayal Umeshwar		& -2975 & 25	& 44 	& 34   	& 306 		& 8553 		& 27.95 	\\ 
Maier David		& -3096 & 26	& 45 	& 32   	& 331 		& 9774 		& 29.53 	\\ 
Wiederhold Gio		& -3320 & 27	& 43 	& 38   	& 315 		& 8376 		& 26.59 	\\ 
Ramakrishnan Raghu		& -4249 & 28	& 52 	& 23   	& 348 		& 11143 		& 32.02 	\\ 
Snodgrass Rick		& -4293 & 29	& 41 	& 42   	& 297 		& 6203 		& 20.89 	\\ 
Srivastava Divesh		& -4333 & 30	& 44 	& 35   	& 317 		& 7679 		& 24.22 	\\ 
Ceri Stefano		& -4355 & 31	& 45 	& 33   	& 345 		& 9145 		& 26.51 	\\ 
Kriegel Hans-Peter		& -5034 & 32	& 46 	& 31   	& 451 		& 13596 		& 30.15 	\\ 
Stonebraker M.		& -5643 & 33	& 62 	& 11   	& 380 		& 14073 		& 37.03 	\\ 
Halevy Alon		& -5858 & 34	& 71 	& 5   	& 392 		& 16933 		& 43.20 	\\ 
Abbadi Amr		& -6906 & 35	& 39 	& 45   	& 361 		& 5652 		& 15.66 	\\ 
Gray Jim		& -7953 & 36	& 54 	& 18   	& 508 		& 16563 		& 32.60 	\\ 
Faloutsos Christos		& -8509 & 37	& 68 	& 7   	& 484 		& 19779 		& 40.87 	\\ 
Jensen Christian		& -8566 & 38	& 44 	& 37   	& 389 		& 6614 		& 17.00 	\\ 
Agrawal Divyakant		& -9199 & 39	& 40 	& 44   	& 433 		& 6521 		& 15.06 	\\ 
Aalst W.		& -9811 & 40	& 48 	& 28   	& 468 		& 10349 		& 22.11 	\\ 
Weikum Gerhard		& -11700 & 41	& 44 	& 36   	& 467 		& 6912 		& 14.80 	\\ 
Sheth Amit		& -12193 & 42	& 58 	& 17   	& 488 		& 12747 		& 26.12 	\\ 
Carey Michael		& -14606 & 43	& 60 	& 14   	& 488 		& 11074 		& 22.69 	\\ 
Franklin Michael		& -14765 & 44	& 60 	& 13   	& 559 		& 15175 		& 27.15 	\\ 
Han Jiawei		& -15410 & 45	& 77 	& 3   	& 653 		& 28942 		& 44.32 	\\ 
Jajodia Sushil		& -15483 & 46	& 53 	& 21   	& 554 		& 11070 		& 19.98 	\\ 
Mylopoulos John		& -15513 & 47	& 53 	& 20   	& 569 		& 11835 		& 20.80 	\\ 
Garcia-molina Hector		& -17423 & 48	& 92 	& 1   	& 605 		& 29773 		& 49.21 	\\ 
Bertino Elisa		& -27058 & 49	& 49 	& 27   	& 805 		& 9986 		& 12.40 	\\ 
Yu Philip		& -27727 & 50	& 63 	& 9   	& 789 		& 18011 		& 22.83 	\\ 

\hline
\end{tabular}
\normalsize
\caption{Rank table by $PT$  of sample "DataBases"}
\label{tab_rank_table_PT_s25}
\end{center}
\end{table}

\begin{table}[!bt]
\begin{center}
\footnotesize

\begin{tabular}{|l@{}||@{\hspace{.1cm}}rr@{\hspace{.1cm}}|@{\hspace{.1cm}}rr@{\hspace{.1cm}}|r|r|r|}\hline
\multirow{2}{*}{\bf Author } 	&\multicolumn{2}{@{\hspace{-.1cm}}|c@{\hspace{.1cm}}}{\bf $PT$} 	&\multicolumn{2}{@{\hspace{-.1cm}}|c@{\hspace{.1cm}}|}{\bf $h$} & \multicolumn{1}{|c|}{\multirow{2}{*}{\bf $p$}}& \multicolumn{1}{|c|}{\multirow{2}{*}{\bf $C$}}& \multicolumn{1}{|c|}{\multirow{2}{*}{\bf $C/p$}}\\
	&val	&pos	&val	&pos &  &  & \\\hline\hline
Donoho David		& 7508 & 1	& 72 	& 2   	& 350 		& 27524 		& 78.64 	\\ 
Cox Ingemar		& 3464 & 2	& 41 	& 15   	& 210 		& 10393 		& 49.49 	\\ 
Simoncelli Eero		& 2619 & 3	& 47 	& 12   	& 227 		& 11079 		& 48.81 	\\ 
Yeo Boon-lock		& 2131 & 4	& 27 	& 44   	& 77 		& 3481 		& 45.21 	\\ 
Rui Yong		& 1745 & 5	& 33 	& 32   	& 168 		& 6200 		& 36.90 	\\ 
Jain Ramesh		& 1637 & 6	& 36 	& 25   	& 243 		& 9089 		& 37.40 	\\ 
Yeung Minerva		& 1490 & 7	& 24 	& 48   	& 66 		& 2498 		& 37.85 	\\ 
Goljan Miroslav		& 1401 & 8	& 28 	& 41   	& 64 		& 2409 		& 37.64 	\\ 
Wiegand Thomas		& 602 & 9	& 32 	& 33   	& 262 		& 7962 		& 30.39 	\\ 
Fridrich Jessica		& 472 & 10	& 27 	& 45   	& 118 		& 2929 		& 24.82 	\\ 
ELAD MICHAEL		& 156 & 11	& 36 	& 26   	& 216 		& 6636 		& 30.72 	\\ 
Naphade Milind		& 136 & 12	& 24 	& 49   	& 106 		& 2104 		& 19.85 	\\ 
Manjunath B.		& -46 & 13	& 39 	& 20   	& 279 		& 9314 		& 33.38 	\\ 
Orchard M.		& -784 & 14	& 34 	& 30   	& 187 		& 4418 		& 23.63 	\\ 
Wu Min		& -1079 & 15	& 27 	& 46   	& 169 		& 2755 		& 16.30 	\\ 
Li Mingjing		& -1180 & 16	& 28 	& 42   	& 150 		& 2236 		& 14.91 	\\ 
Zhang Ya-Qin		& -2205 & 17	& 36 	& 28   	& 236 		& 4995 		& 21.17 	\\ 
Hauptmann Alexander		& -3183 & 18	& 34 	& 31   	& 243 		& 3923 		& 16.14 	\\ 
Smith John		& -3277 & 19	& 40 	& 17   	& 282 		& 6403 		& 22.71 	\\ 
Zakhor Avideh		& -3468 & 20	& 38 	& 23   	& 268 		& 5272 		& 19.67 	\\ 
Ebrahimi Touradj		& -3520 & 21	& 31 	& 37   	& 272 		& 3951 		& 14.53 	\\ 
MEMON NASIR		& -3736 & 22	& 32 	& 34   	& 286 		& 4392 		& 15.36 	\\ 
Li Shipeng		& -3925 & 23	& 26 	& 47   	& 271 		& 2445 		& 9.02 	\\ 
Hua Xian-sheng		& -4252 & 24	& 24 	& 50   	& 285 		& 2012 		& 7.06 	\\ 
Ma Wei-ying		& -4292 & 25	& 46 	& 13   	& 335 		& 9002 		& 26.87 	\\ 
Ortega Antonio		& -4894 & 26	& 31 	& 36   	& 330 		& 4375 		& 13.26 	\\ 
Xiong Zixiang		& -4950 & 27	& 35 	& 29   	& 308 		& 4605 		& 14.95 	\\ 
Bouman C		& -5592 & 28	& 27 	& 43   	& 380 		& 3939 		& 10.37 	\\ 
Wu Xiaolin		& -6332 & 29	& 31 	& 39   	& 337 		& 3154 		& 9.36 	\\ 
Bovik Alan		& -8008 & 30	& 39 	& 19   	& 507 		& 10244 		& 20.21 	\\ 
Ramchandran Kannan		& -8111 & 31	& 49 	& 11   	& 421 		& 10117 		& 24.03 	\\ 
Liu Bede		& -8784 & 32	& 38 	& 22   	& 436 		& 6340 		& 14.54 	\\ 
Strintzis M.		& -8871 & 33	& 29 	& 40   	& 454 		& 3454 		& 7.61 	\\ 
Chang Edward		& -9099 & 34	& 31 	& 38   	& 448 		& 3828 		& 8.54 	\\ 
Delp Edward		& -10001 & 35	& 37 	& 24   	& 438 		& 4836 		& 11.04 	\\ 
Chen Liang-Gee		& -10311 & 36	& 32 	& 35   	& 478 		& 3961 		& 8.29 	\\ 
Tekalp A.		& -10552 & 37	& 40 	& 18   	& 448 		& 5768 		& 12.88 	\\ 
Unser Michael		& -10801 & 38	& 54 	& 6   	& 465 		& 11393 		& 24.50 	\\ 
Vetterli M.		& -11139 & 39	& 63 	& 4   	& 547 		& 19353 		& 35.38 	\\ 
Jain Anil		& -11474 & 40	& 81 	& 1   	& 590 		& 29755 		& 50.43 	\\ 
Katsaggelos Aggelos		& -11662 & 41	& 36 	& 27   	& 504 		& 5186 		& 10.29 	\\ 
Wang Yao		& -11705 & 42	& 39 	& 21   	& 484 		& 5650 		& 11.67 	\\ 
Chang Shih-Fu		& -11941 & 43	& 52 	& 7   	& 507 		& 11719 		& 23.11 	\\ 
Nahrstedt Klara		& -12286 & 44	& 52 	& 9   	& 492 		& 10594 		& 21.53 	\\ 
Girod Bernd		& -13613 & 45	& 52 	& 8   	& 529 		& 11191 		& 21.16 	\\ 
Pitas Ioannis		& -13849 & 46	& 44 	& 14   	& 515 		& 6875 		& 13.35 	\\ 
Chellappa Rama		& -14092 & 47	& 50 	& 10   	& 604 		& 13608 		& 22.53 	\\ 
Zhang Hongjiang		& -15112 & 48	& 63 	& 5   	& 556 		& 15947 		& 28.68 	\\ 
Kuo C.		& -36848 & 49	& 40 	& 16   	& 1148 		& 7472 		& 6.51 	\\ 
Huang Thomas		& -54047 & 50	& 67 	& 3   	& 1172 		& 19988 		& 17.05 	\\ 

\hline
\end{tabular}
\normalsize
\caption{Rank table by $PT$  of sample "Multimedia"}
\label{tab_rank_table_PT_s35}
\end{center}
\end{table}

\begin{table}[!bt]
\begin{center}
\footnotesize

\begin{tabular}{|l@{}||@{\hspace{.1cm}}rr@{\hspace{.1cm}}|@{\hspace{.1cm}}rr@{\hspace{.1cm}}|r|r|r|}\hline
\multirow{2}{*}{\bf Author } 	&\multicolumn{2}{@{\hspace{-.1cm}}|c@{\hspace{.1cm}}}{\bf $PT$} 	&\multicolumn{2}{@{\hspace{-.1cm}}|c@{\hspace{.1cm}}|}{\bf $h$} & \multicolumn{1}{|c|}{\multirow{2}{*}{\bf $p$}}& \multicolumn{1}{|c|}{\multirow{2}{*}{\bf $C$}}& \multicolumn{1}{|c|}{\multirow{2}{*}{\bf $C/p$}}\\
	&val	&pos	&val	&pos &  &  & \\\hline\hline
Jacobson Van		& 19982 & 1	& 44 	& 44   	& 161 		& 25130 		& 156.09 	\\ 
Floyd Sally		& 18059 & 2	& 66 	& 9   	& 222 		& 28355 		& 127.73 	\\ 
Balakrishnan Hari		& 14444 & 3	& 72 	& 7   	& 272 		& 28844 		& 106.04 	\\ 
Johnson David		& 12180 & 4	& 54 	& 21   	& 263 		& 23466 		& 89.22 	\\ 
Morris Robert		& 11685 & 5	& 57 	& 16   	& 305 		& 25821 		& 84.66 	\\ 
Handley M.		& 10763 & 6	& 47 	& 35   	& 201 		& 18001 		& 89.56 	\\ 
Perkins C.		& 9609 & 7	& 52 	& 25   	& 373 		& 26301 		& 70.51 	\\ 
Paxson Vern		& 8871 & 8	& 60 	& 12   	& 233 		& 19251 		& 82.62 	\\ 
Stoica Ion		& 8558 & 9	& 63 	& 11   	& 266 		& 21347 		& 80.25 	\\ 
Heidemann John		& 8059 & 10	& 47 	& 36   	& 237 		& 16989 		& 71.68 	\\ 
Culler David		& 7552 & 11	& 84 	& 3   	& 386 		& 32920 		& 85.28 	\\ 
Shenker Scott		& 5754 & 12	& 97 	& 1   	& 508 		& 45621 		& 89.81 	\\ 
Govindan Ramesh		& 5356 & 13	& 55 	& 19   	& 287 		& 18116 		& 63.12 	\\ 
Estrin Deborah		& 5348 & 14	& 90 	& 2   	& 479 		& 40358 		& 84.25 	\\ 
Crovella Mark		& 4886 & 15	& 46 	& 38   	& 172 		& 10682 		& 62.10 	\\ 
Perrig Adrian		& 4304 & 16	& 58 	& 14   	& 247 		& 15266 		& 61.81 	\\ 
Lu Songwu		& 3430 & 17	& 44 	& 45   	& 129 		& 7170 		& 55.58 	\\ 
Akyildiz Ian		& 3089 & 18	& 53 	& 23   	& 401 		& 21533 		& 53.70 	\\ 
Kleinrock Leonard		& 1986 & 19	& 51 	& 31   	& 282 		& 13767 		& 48.82 	\\ 
Knightly Edward		& 263 & 20	& 41 	& 50   	& 172 		& 5634 		& 32.76 	\\ 
Peterson L.		& -652 & 21	& 54 	& 22   	& 292 		& 12200 		& 41.78 	\\ 
Hubaux Jean-Pierre		& -653 & 22	& 45 	& 43   	& 247 		& 8437 		& 34.16 	\\ 
Vaidya Nitin		& -1242 & 23	& 50 	& 32   	& 337 		& 13108 		& 38.90 	\\ 
Zhang Lixia		& -1609 & 24	& 55 	& 20   	& 374 		& 15936 		& 42.61 	\\ 
Low Steven		& -1796 & 25	& 45 	& 42   	& 291 		& 9274 		& 31.87 	\\ 
Boudec Jean-Yves		& -2338 & 26	& 44 	& 46   	& 258 		& 7078 		& 27.43 	\\ 
Win Moe		& -2619 & 27	& 46 	& 37   	& 341 		& 10951 		& 32.11 	\\ 
Rexford Jennifer		& -2632 & 28	& 49 	& 33   	& 269 		& 8148 		& 30.29 	\\ 
Zhang Hui		& -3344 & 29	& 52 	& 27   	& 352 		& 12256 		& 34.82 	\\ 
Srikant R.		& -3827 & 30	& 46 	& 39   	& 328 		& 9145 		& 27.88 	\\ 
Diot Christophe		& -4054 & 31	& 52 	& 30   	& 290 		& 8322 		& 28.70 	\\ 
Simon Marvin		& -4450 & 32	& 42 	& 49   	& 370 		& 9326 		& 25.21 	\\ 
Ammar Mostafa		& -4547 & 33	& 43 	& 48   	& 308 		& 6848 		& 22.23 	\\ 
Kurose Jim		& -5114 & 34	& 59 	& 13   	& 391 		& 14474 		& 37.02 	\\ 
Campbell Andrew		& -6036 & 35	& 46 	& 41   	& 348 		& 7856 		& 22.57 	\\ 
Chlamtac I.		& -6274 & 36	& 43 	& 47   	& 357 		& 7228 		& 20.25 	\\ 
Crowcroft Jon		& -6863 & 37	& 48 	& 34   	& 404 		& 10225 		& 25.31 	\\ 
WHITT W.		& -7759 & 38	& 52 	& 29   	& 394 		& 10025 		& 25.44 	\\ 
Goldsmith A.		& -7819 & 39	& 57 	& 17   	& 479 		& 16235 		& 33.89 	\\ 
Srivastava Mani		& -8139 & 40	& 57 	& 18   	& 423 		& 12723 		& 30.08 	\\ 
Paulraj A.		& -8421 & 41	& 64 	& 10   	& 442 		& 15771 		& 35.68 	\\ 
Schulzrinne Henning		& -11050 & 42	& 53 	& 24   	& 555 		& 15556 		& 28.03 	\\ 
Garcia-Luna-Aceves J.		& -11169 & 43	& 46 	& 40   	& 460 		& 7875 		& 17.12 	\\ 
Nahrstedt Klara		& -12286 & 44	& 52 	& 28   	& 492 		& 10594 		& 21.53 	\\ 
Cioffi J.		& -14685 & 45	& 52 	& 26   	& 575 		& 12511 		& 21.76 	\\ 
Mukherjee B.		& -15702 & 46	& 58 	& 15   	& 535 		& 11964 		& 22.36 	\\ 
Katz Randy		& -27820 & 47	& 78 	& 5   	& 757 		& 25142 		& 33.21 	\\ 
Towsley Don		& -31929 & 48	& 82 	& 4   	& 793 		& 26373 		& 33.26 	\\ 
GERLA MARIO		& -37464 & 49	& 67 	& 8   	& 945 		& 21362 		& 22.61 	\\ 
Giannakis Georgios		& -44707 & 50	& 77 	& 6   	& 932 		& 21128 		& 22.67 	\\ 

\hline
\end{tabular}
\normalsize
\caption{Rank table by $PT$  of sample "Networks"}
\label{tab_rank_table_PT_s45}
\end{center}
\end{table}

Tables \ref{tab_rank_table_PT_s25},\ref{tab_rank_table_PT_s35} and \ref{tab_rank_table_PT_s45} resent the top 50 researchers from each of the areas of Databases (Table \ref{tab_rank_table_PT_s25}), Multimedia (Table \ref{tab_rank_table_PT_s35}) and Networks (Table \ref{tab_rank_table_PT_s45}). We have re-rank then by using the $PT$ metric.

\section{Conclusions \& Future Work}
\label{sec:conclusion}

In this article we have defined two new areas on an author's citation curve:
\begin{itemize}
\item 
the {\it tail complement penalty area} (TC-area), i.e. the complement of the tail with respect to the rectangle $h \times (p-h)$, with size $C_{TC}$.
\item
the {\it ideal complement penalty area} (IC-area), i.e. a complement with respect to the square $p \times p$, with size $C_{IC}$.
\end{itemize}
By using the above areas we have defined two new metrics:
\begin{itemize}
\item
the {\it penalty index based on the TC-area}, in short $PT$ index, and
\item
the {\it penalty index based on the IC-area}, in short $PI$ index.
\end{itemize}
We have performed several experiments to study the behavior of the $PT$ and $PI$ indices. 
For this purpose, we have generated 3 datasets (with random authors, with prolific authors and with authors with high \hi) by extracting data from the Microsoft Academic Search database. 
Our contribution is threefold:
\begin{itemize}
\item
we have shown that both new indices, $PT$ and $PI$, are un-correlated to previous ones, such as the \hi. 
\item
we have used these new indices, in particular $PT$, to rank authors in general and, in particular, to split the population of authors into two distinct groups: the "influential" ones with $PT>0$ vs. the "mass producers" with $PT<0$. Finally,
\item
it has been shown that ranking authors with the $PT$ metric is more robust than the \hi with respect to noise of self-citations.
\end{itemize}
A future work will be integrate the notion of \chi \cite{SidiropoulosKM07} into the $PT$ metric. 

\vspace*{-0.5\baselineskip}
\bibliographystyle{plain}
\bibliography{Hindexareas}
\end{document}